\documentclass[12pt]{article}
\usepackage{amssymb}
\usepackage{latexsym}
\newtheorem{proposition}{Proposition}
\newtheorem{corollary}{Corollary}
\newtheorem{theorem}{Theorem}

\begin{document}
\begin{flushright} {\today}
\end{flushright}
\vspace*{1cm}

\begin{center} {\large {\bf Asymptotic Behavior in Polarized and
Half-Polarized $U(1)$ Symmetric Vacuum Spacetimes}}

\vskip .75cm {\bf James Isenberg}
\vskip.5cm {Department of Mathematics and Institute of
Theoretical Science\\ University of Oregon, Eugene, OR
97403-5203, USA} \vskip .75cm {\bf Vincent Moncrief} \vskip.5cm
{Department of Physics and Department of Mathematics\\ Yale
University\\ New Haven, CT 06520,USA} \end{center}

\begin{abstract} We use the Fuchsian algorithm to study the
behavior near the singularity of certain families of $U(1)$
Symmetric solutions of the vacuum Einstein equations (with the
$U(1)$ isometry group acting spatially). We consider an analytic
family of polarized solutions with the maximum number of
arbitrary functions consistent with the polarization condition
(one of the ``gravitational degrees of freedom'' is turned off)
and show that all members of this family are asymptotically
velocity term dominated (AVTD) as one approaches the singularity.
We show that the same AVTD behavior holds for a family of ``half
polarized'' solutions, which is defined by adding one extra
arbitrary function to those characterizing the polarized
solutions. (The full set of non-polarized solutions involves two
extra arbitrary functions). We begin to address the issue of
whether AVTD behavior is independent of the choice of time
foliation by showing that indeed AVTD behavior is seen for a wide
class of choices of harmonic time in the polarized and
half-polarized ($U(1)$ Symmetric vacuum) solutions discussed
here. \end{abstract}
\newpage

\section{Introduction}

During the last few years, our understanding of the behavior of
cosmological solutions\footnote{Here we follow Bartnik's
convention \cite{B88} of calling a spacetime a ``cosmological
solution'' if, in addition to being a solution of Einstein's
equations, it is globally hyperbolic, with compact Cauchy
surfaces.} near their big bang singularities has increased
significantly. There is more and more evidence, both numerical
\cite{BM93} \cite{BG98} \cite{WIB98} \cite{BIW01} \cite{BM98a}
\cite {BM98b} and analytical \cite{KR98} \cite{IK99} \cite{Re00}
\cite{AR01}, that wider and wider families of such spacetimes
exhibit either asymptotically velocity term dominated (``AVTD'')
behavior, or Mixmaster behavior, in a neighborhood of their
singularities.

A solution has AVTD behavior if the metric (when expressed in
suitable gauge) asymptotically behaves at each spatial point like
a Kasner spacetime metric, with the Kasner parameters generally
varying from point to point\footnote{See Section 3 for a more
precise definition of AVTD behavior.}. A solution exhibits
Mixmaster behavior if, instead of asymptotic Kasner evolution at
each point, one asymptotically has at each point the oscillatory
behavior of bouncing from one Kasner epoch to another, as seen in
Bianchi IX cosmologies. The idea that a generic solution should
have one or the other of these behaviors was first suggested by
Belinskii, Khalatnikov, and Lifschitz \cite{BKL70,BKL82},
and is therefore known as the ``BKL conjecture.''

For spacetimes with a $U(1)$ isometry group and $T^3$ spatial
topology, there is strong numerical support for the BKL
conjecture. Extensive numerical studies of these spacetimes
indicate that if the Killing field generating the isometry is
hypersurface orthogonal (these are the so-called ``polarized
solutions,'' since one of the two gravitational degrees of
freedom is essentially turned off), then AVTD behavior is seen
\cite{BM98a}; while for generic $U(1)$ Symmetric spacetimes with
$T^3$ spatial topology, one finds Mixmaster behavior
\cite{BM98b}. These results agree with the conclusions of earlier
studies of $U(1)$ Symmetric spacetimes \cite{GM93}, which used
formal expansions and heuristic analysis to argue that the BKL
conjecture likely holds for them.

In this work, we rigorously prove some of these results, using
Fuchsian methods. More specifically, we show that for rather
general sets of analytic data for polarized $U(1)$ Symmetric
solutions, the solutions are AVTD. Furthermore, we identify a
class of $U(1)$ Symmetric solutions, intermediate between the
polarized and the generic classes, which is AVTD as well. The
solutions in this class, which we call ``half-polarized,'' are
intermediate in the sense that, modulo gauge freedom and the
constraint equations, the polarized solutions are characterized
by two arbitrary functions on
$T^2$, the half-polarized by three such functions, and the
nonpolarized by four. So in a rough sense, what we find is a
distinct way to turn on half of the second degree of freedom at
each point, without removing the AVTD behavior. We further show
that the class of $U(1)$ Symmetric solutions with AVTD behavior
includes spacetimes which are neither polarized nor
half-polarized, since we can find such solutions by applying
$SL(2,R)$ parametrized Geroch transformations to the
half-polarized solutions and still retain AVTD behavior.

Fuchsian methods have become an important tool in recent years
for proving the existence of AVTD behavior in cosmological
solutions of Einstein's equations. Kichenassamy and Rendall first
used these techniques, in 1998, to establish AVTD behavior in
(unpolarized) Gowdy spacetimes \cite{KR98}. Subsequent work
\cite{IK99} showed that in polarized $T^2$ Symmetric solutions
with nonvanishing twist (the vanishing of the twist characterizes
the Gowdy solutions as a subfamily of the $T^2$ Symmetric
cosmological spacetimes), AVTD behavior is found. Especially
interesting is the very recent result \cite{AR01} which uses
Fuchsian methods to show that AVTD behavior occurs in
cosmological solutions (spatially $T^3$) of the Einstein-scalar
field equations with {\it no} assumed symmetries.

In none of these works, including the work discussed here, do the
Fuchsian methods show that {\it all} solutions in the family
under consideration (general Gowdy, polarized $T^2$ Symmetric,
polarized and half-polarized $U(1)$ Symmetric, etc) necessarily
have AVTD behavior. Rather, the Fuchsian methods show that in
each family of solutions, there is an analytic subfamily which is
defined by the same number of free functions as the full family,
and whose members all exhibit AVTD behavior. Thus, in a rough
sense, Fuchsian methods show that AVTD behavior is stable,
occurring in ``open subsets'' of the given full family. It is
expected, based on numerical evidence, that AVTD behavior is
generic in these families; however, this has not been proven yet
in any family except for the polarized Gowdy solutions.

The polarized and half-polarized $U(1)$ Symmetric spacetimes with
which we work here comprise the least restricted family of vacuum
solutions in which AVTD behavior has been proven to exist. It is
useful to note that the recent work showing that AVTD behavior
occurs in Einstein-scalar field solutions with no symmetry does
not imply our $U(1)$ results. Indeed heuristic arguments first
presented by Belinskii, Khalatnikov, and Lifschitz \cite{BKL82}
indicate that the presence of a nonvanishing scalar field is
likely to remove the inevitability of the ``potential bounces''
which lead to Mixmaster behavior, with AVTD behavior resulting.
It is expected that vacuum solutions with no isometries are
generically not AVTD.

It would, of course, be very nice if Fuchsian type methods could
be used to rigorously prove that Mixmaster behavior occurs in
those families of solutions (like the magnetic Gowdy spacetimes
\cite{WIB98}, the nonpolarized $T^2$ Symmetric spacetimes
\cite{BIW01}, and nonpolarized $U(1)$ Symmetric spacetimes
\cite{BM98b}) in which Mixmaster behavior has been observed
numerically. However, it is not known how to do this. Indeed,
only very recently has Mixmaster behavior been rigorously
verified in spatially homogeneous spacetimes \cite{W00}
\cite{Ri01}.

In all Fuchsian studies of cosmological solutions prior to ours,
the choice of spacetime foliation has been more or less rigidly
determined by the family of spacetimes and the analysis. For
example, for the Gowdy and for the $T^2$ Symmetric spacetimes,
the analysis of solutions is simplest if one uses the ``areal''
(or ``Gowdy'') foliation \cite{Go74} \cite{BCIM97}, in which the
$t = constant$ hypersurfaces each consist of $T^2$ orbits of
constant area; this foliation is (up to scaling) unique. In our
present study of polarized
$U(1)$ Symmetric spacetimes, however, the analysis is carried out
using a harmonic time foliation, as detailed below. Harmonic time
is not unique, and so we have the opportunity here to consider the
issue of whether, and in what sense, observed AVTD behavior
depends on the choice of foliation. We show that if, in a fixed
$U(1)$ Symmetric solution $(T^3\times R,g)$, AVTD behavior is
seen using one choice of harmonic time, then there is a full (two
free functions on $T^2$) family of other choices of harmonic time
for $(T^3\times R,g)$ such that AVTD behavior is seen using each
of them as well. We also discuss the extent to which we expect
AVTD behavior to be seen using other foliations.

The bulk of this paper is devoted to the verification, using
Fuchsian methods, that AVTD behavior occurs in polarized and
half-polarized $U(1)$ Symmetric vacuum solutions on $T^3 \times
R$. We begin by reviewing (in Section 2), the form of the metric
and the form of the field equations we shall work with here. We
obtain these forms by starting with the generic metric for $U(1)$
Symmetric spacetimes, and then imposing certain gauge conditions,
and certain restrictions on the fields which are preserved by the
field equations. We write the field equations in canonical,
Hamiltonian form. We note in Section 2 that in setting one of the
fields together with its conjugate momentum to zero, we define
the polarized solutions. We further note that the half-polarized
solutions cannot be characterized by restrictions on the fields
at an arbitrary finite time; rather, they are characterized by
certain asymptotic conditions near the singularity.

In Section 3, we write the asymptotic (near the singularity)
ansatz for the polarized solutions, and then for the
half-polarized solutions. The asymptotic ansatz for the polarized
fields (forgetting the constraints) involves eight free functions
(the ``asymptotic data'') on $T^2$ together with eight remainder
functions depending on time as well as on $T^2$. (There are eight
rather than two because the constraint equations have not been
imposed, and some gauge freedom remains.) Substituting this
ansatz into the Einstein evolution equations, one obtains
evolution equations for these remainder functions (with the free
functions on $T^2$ appearing as parameters). We show in Section 3
that for all choices of the asymptotic data which satisfy certain
linear inequalities, these evolution equations take the special
Fuchsian form. It then follows from a result of Kichenassamy and
Rendall \cite{KR98}, that for any such choice of the asymptotic
data, there is a unique solution for the remainder functions,
decaying to zero near the singularity. This behavior guarantees
that solutions in the (polarized) ansatz form are AVTD. The
analysis for solutions in the half-polarized asymptotic ansatz
form is similar, with similar conclusions.

The discussion in Section 3 ignores the constraint equations. In
Section 4, we show that if the asymptotic data in the asymptotic
ansatz expressions satisfy certain asymptotic constraints
$(\ref{70})-(\ref{72})$, then the corresponding solution
satisfies the Einstein constraints. The converse is true as well;
so the Einstein constraints are (at least among solutions of the
evolution equations) equivalent to these asymptotic constraints
$(\ref{70})-(\ref{72})$. It follows that the appropriate sets of
free functions, restricted to satisfy the asymptotic constraints,
parameterize the corresponding families of AVTD polarized and
half-polarized solutions.

The work done in Sections 2-4 presumes a fixed choice of harmonic
time foliation. In Section 5, we discuss the relation between
alternative harmonic time foliations, and we show that one sees
AVTD behavior using any one of them. The asymptotic relation
between alternative harmonic time foliations explains why this
makes sense, and hints at criteria for predicting whether one
sees AVTD behavior using other choices of time foliation and
observers.

We show in Section 6, using the $SL(2,R)$ Geroch transformation
\cite{Ge71}, that there are $U(1)$ Symmetric solutions, beyond
those discussed here, in which AVTD behavior is seen. We
speculate on further classes of solutions which exhibit AVTD
behavior, and make concluding remarks, in Section 7.

\section{$U(1)$ Symmetric Vacuum Spacetimes}

If we choose $2\pi$-periodic coordinates $(x^a, x^3) = (x^1, x^2,
x^3)$ on $T^3$, with the Killing field expressed as $\partial
\over \partial x^3$, then the generic $U(1)$ Symmetric spacetime
metric on $T^3 \times R^1$ may be written in the form
\begin{eqnarray}
g= e^{-2\phi} \left[-N^2 e^{-4\tau} d\tau^2 +
e^{-2\tau} e^\Lambda e_{ab} \, dx^a dx^b\right] + e^{2\phi}(dx^3
+ \beta_a dx^a)^2\quad,
\label{1}
\end{eqnarray}
with the functions $\phi$, $N$, and $\Lambda$, and the form field
$\beta_a$, and the unit determinant symmetric tensor field
$e_{ab}$, all functions of $x^a$ and the time coordinate $\tau$
(independent of $x^3$). Here the latin indices $a$, $b$ take the
values 1, 2; note that we shall find it convenient to write
$(x^1, x^2) = (u, v)$.

The metric form (\ref{1}) does involve some restrictions on the
gauge freedom, in that we have aligned the Killing field with
$x^3$ for all time, and we have set the corresponding shift
vector to zero. We now further restrict the gauge freedom by
setting the lapse function $N$ equal to $e^{\Lambda}$. This
choice of the lapse does not restrict the initial choice of
Cauchy slice; however, once a first slice is chosen, all others
are determined. We note that it follows from the condition $N =
e^{\Lambda}$ that the time function $\tau$ satisfies the equation
\begin{eqnarray}
\Box \tau = 0
\label{2}
\end{eqnarray}
where $\Box$ is the wave operator corresponding to
the metric $g$.\footnote{One has $\Box \tau = 0$ for $\Box$
corresponding to the $2 + 1$ metric $\tilde{g} = -N^2 e^{-4\tau}
d\tau^2 + e^{-2\tau} e^\Lambda e_{ab} \, dx^a dx^b$ as well as
for $\Box$ corresponding to the $3 + 1$ metric in equation
(\ref{1}).} Hence this choice of slicing is sometimes called a
``harmonic time.''\footnote{This terminology is a consequence of
the somewhat loose practice of using the word ``harmonic'' to
refer to solutions of the equation
$\Box \tau = 0$ as well as to the equation $\nabla^2 \tau = 0$.}

Before writing down the full system of Einstein's equations for
the $U(1)$ symmetric spacetimes in terms of the metric (\ref{1}),
we wish to make a small restriction which considerably simplifies
the analysis. To state this condition, we first write out the
projection of the Einstein supermomentum constraint along the
Killing field $\partial \over \partial x^3$; it takes the form
\footnote{The form of this constraint is reminiscent of the Gauss
law constraint for electromagnetism. This reflects the fact that
one may view 3 + 1 vacuum gravity with $U(1)$ symmetry as a
Kaluza-Klein theory for 2 + 1 Einstein-Maxwell fields, plus a
Jordan scalar field.}
\begin{eqnarray}
f^a ,_a = 0 \label{3}
\end{eqnarray}
where $f^a$ is the momentum conjugate to
$\beta_a$. The general solution to (\ref{3}) can be written as
\begin{eqnarray}
f^a = \epsilon^{ab}w, _b + h^a
\label{4}
\end{eqnarray}
where $w$ is a scalar function,
$\epsilon^{ab}:=\left(\begin{array}{cc}
0&1\\-1&0 \end{array}\right)$, and $h^a$ is dual to a harmonic
one-form on $T^2$. For the $T^3 \times R$ spacetimes of interest
here\footnote{We note that if one studies $U(1)$ Symmetric
spacetimes on $\left(\Sigma ^2 \times S^1\right) \times R $, for
more general surfaces $\Sigma ^2$, the same equation (\ref{3})
appears, and one has the same sort of solution (\ref{4}).
However, for $\Sigma ^2 = S^2, (h^a)$ necessarily vanishes, while
for higher genus $\Sigma ^2, (h^a)$ can be more complicated.} one
finds (using the rest of Einstein's equations), that
$(h^a)=\left(\begin{array}{c} C^1\\C^2\end{array} \right)$, for
$C^1$ and $C^2$ constant in time as well as in space.

So far, no restrictive assumptions have been made. Now, however,
we set both $C^1$ and $C^2$ to zero. Correspondingly, we write
$\beta_a$ in terms of an integral involving one free function,
which we call $r$ (see \cite{M86} \cite{BM98a} for details), and
it follows that the canonical pair $\left(\beta_a,f^a\right)$ is
replaced by the pair $\left(w,r \right)$. These restrictions are
consistent with Einstein's equations; they are somewhat analogous
to the vanishing of the two ``twist constants'' which defines the
Gowdy spacetimes as a subfamily of the $T^2$ Symmetric
spacetimes. One might wish elsewhere to consider what happens if
$C^1$ or $C^2$ is nonzero, but we shall not pursue the issue
here.

If we now choose the following convenient (and {\it
non}-restrictive) form for the unit determinant tensor $e_{ab}$,
\begin{eqnarray}
e_{ab} = {1 \over 2}\left(\begin{array}{ll}
e^{2z} + e^{-2z} (1 + x)^2&e^{2z} + e^{-2z} (x^2 - 1)\\e^{2z} +
e^{-2z} (x^2 - 1)&e^{2z} + e^{-2z} (1 - x)^2\end{array} \right)
\label{5}
\end{eqnarray}
(where $z$ and $x$ are functions of $u$,
$v$ and $\tau$), then the Einstein evolution equations for these
spacetimes can be expressed as a Hamiltonian system for the
fields $\left\{\left.\phi, \Lambda, w, x, z\right.\right)$ and
their respective conjugate momenta $\left\{ p, p_{\Lambda}r,
p_{x}, p_z\right\}$. The Hamiltonian for this system is
\begin{eqnarray}
H = \int_{T^2} h\, du dv
\label{6a}
\end{eqnarray}
with
\begin{eqnarray}
h= \hspace{-.6cm}&&{1 \over
8} p^2_z + {1 \over 2} e^{4z}p^2_{x} + {1 \over 8} p^2 + {1
\over 2} e^{4\phi}r^2 - {1 \over 2}p^2_{\Lambda} + 2
p_{\Lambda}\nonumber\\ && +
e^{-2\tau}\left\{\left(e^{\Lambda}e^{ab} \right),_{ab} -
\left(e^{\Lambda}e^{ab} \right),_a \Lambda,_b +
e^{\Lambda}\left[\left(e^{-2z} \right),_u x, _v - \left(e^{-2z}
\right),_vx,_u\right]\right.\nonumber\\ &&\left. +
2e^{\Lambda}e^{ab}\phi,_a \phi,_b + {1 \over 2}
e^{\Lambda}e^{-4\phi}e^{ab}w,_a w,_b\right\}
\label{6b}
\end{eqnarray}
where $e^{ab}$ is the matrix inverse of $e_{ab}$
(see equation (\ref{5})). The evolution equations for
$\left\{\phi, \Lambda, w, x, z; p, p_{\Lambda}, r, p_{x}, p_z
\right\}$ are obtained via Hamilton's equations, using this
function $H$.

There are also constraint equations which this data must satisfy.
They take the form
\begin{eqnarray}
0&=& {\cal H}_0\nonumber\\ &=& h - 2p_{\Lambda}
\label{7a}
\end{eqnarray}
\begin{eqnarray}
0= \hspace{-.6cm}&&{\cal H}_u\nonumber\\ =\hspace{-.6cm}&&p_zz,_u
+ p_{x}x,_u + p_{\Lambda} \Lambda, _u -p_{\Lambda},_u + p\phi,_u
+ rw,_u\nonumber\\ &&+{1 \over 2}\left\{\left[e^{4z}- (1 + x)^2
\right]p_{x} - (1 + x) p_z \right\},_v\nonumber\\ &&-{1 \over
2}\left\{\left[e^{4z}+ \left(1 - x^2\right) \right]p_{x} - x p_z
\right\},_u
\label{7b}
\end{eqnarray}
\begin{eqnarray}
0 =\hspace{-.6cm}&& {\cal H}_v\nonumber\\
=\hspace{-.6cm}&&p_zz,_v + p_{x}x,_v + p_{\Lambda} \Lambda, _v
-p_{\Lambda},_v + p\phi,_v + rw,_v\nonumber\\ &&-{1 \over
2}\left\{\left[e^{4z} - \left(1 - x\right)^2 \right]p_{x} + (1 -
x) p_z\right\},_u \nonumber\\ &&+{1 \over
2}\left\{\left[e^{4z}+\left(1 - x^2\right) \right]p_{x} - x p_z
\right\},_v
\label{7c}
\end{eqnarray}
The first of these is the
superHamiltonian constraint. The others are supermomentum
constraints. Recall that one of the supermomentum constraints
--equation (\ref{3})--has been solved, and is therefore of no
further interest. For later purposes, it is important to note
that ${\cal H}_0$ is written as a double density, while ${\cal
H}_u$ and ${\cal H}_v$ are expressed as single densities. This
fact is irrelevant in seeking solutions to the constraints, but
it is important in verifying the preservation of the constraints
under evolution (see Section 4).

In an appropriate sense, the Hamiltonian system
(\ref{6a})-(\ref{6b}) with constraints (\ref{7a})-(\ref{7c}) (and
with the remaining gauge freedom described above) has two
gravitational degrees of freedom. If one can set one of the
functions $\phi$ or $w$, together with the corresponding
conjugate momentum $p$ or $r$, to zero on some Cauchy surface,
and if the evolution equations preserve the vanishing of these
two conjugate variables, then one has reduced the system to one
gravitational degree of freedom, and the resulting family of
solutions is said to be ``polarized.'' Inspecting the Hamiltonian
(\ref{6a})-(\ref{6b}), one finds that the only conjugate pair for
which this can be done is $(w,r)$. Thus, the polarized $U(1)$
Symmetric spacetimes are those with the canonical variables
$\left\{\phi, \Lambda, x, z; p, p_{\Lambda}, p_{x}, p_z
\right\}$, with the Hamiltonian given by (\ref{6a})-(\ref{6b})
with $w = 0$ and $r = 0$, and with the constraints given by
(\ref{7a})-(\ref{7c}) with $w = 0$ and $r = 0$. We note that for
these polarized solutions, the spacetime metric takes the form
(\ref{1}), with $\beta_a = 0$.

As noted in the introduction, we find AVTD behavior in a family
of ``half-polarized'' $U(1)$ Symmetric solutions, as well as in
the family of polarized $U(1)$ Symmetric solutions as just
described. For these half-polarized spacetimes, rather than
setting a conjugate pair of variables to zero, one ties the
conjugate pair together, so that effectively the pair involves
one free function instead of two. It is not known whether this
can be done consistently in terms of initial data on a Cauchy
surface. It can, however, be done in terms of asymptotic data, as
we show in the next section.

\section{Fuchsian Study of the Evolution Equations}

While AVTD behavior can be described in terms of asymptotic
approach to Kasner evolution of the metric fields at each spatial
point (see Introduction), for the purposes of proof, the
following formulation is more useful.

\subsubsection*{Definition (AVTD Behavior)}

{\it A cosmological solution $(M, g)$ exhibits AVTD behavior if
there exists a global spacelike foliation $\Sigma_t$ of $(M, g)$,
and there exists a solution
$(M, \hat{g})$ of the Einstein VTD equations (which are obtained
by dropping all spatial derivatives in the evolution equations
written with respect to $\Sigma_t$, and dropping all spatial
derivatives in the super Hamiltonian constraint) such that if
$t_s \in [-\infty, +\infty]$ labels the singularity in $(M, g)$,
then\footnote{Although this definition in principle makes sense
for any choice of norm $|\,|$, for our purposes here, we presume
that $|\,|$ is the absolute value norm for each of the components
of
$g$, evaluated independently for each spatial point, so the
convergence is pointwise, $L^{\infty}$.} ${\lim}_{t \rightarrow
t_s}\left|g(t) - \hat{g}(t) \right| = 0$.}

Thus, to show that the spacetimes in a given family of solutions
have AVTD behavior, one first needs to find an appropriate
foliation, with the singularity occurring at a well-defined value
of the time parameter. While it is often useful in Fuchsian
analyses to choose coordinates so that the singularity occurs at
$t = 0$, for the $U(1)$ Symmetric spacetimes, it is convenient to
work with the time coordinate $\tau$ (introduced in Section 2),
in terms of which the singularity occurs at $\tau \rightarrow
\infty$. Note that in terms of $t = e^{-\tau}$, the singularity
occurs at $t = 0$; however the use of $\tau$ instead of t
simplifies the analysis here.

We next identify the VTD equations for the $U(1)$ Symmetric
spacetimes. The VTD evolution equations\footnote{In this section,
we ignore the constraint equations, both for the VTD system and
for the full Einstein system.}, written with respect to $\tau$
and the coordinates chosen in Section 2, can be obtained by
varying the Hamiltonian
\begin{eqnarray}
\hat{H} = \int
_{T^2}\left[{1 \over 8}p^2_z + {1 \over 2}e^{4z} p^2_x + {1 \over
8} p^2 + {1 \over 2}e^{4\phi} r^2 - {1 \over 2}p^2_{\Lambda} +
2p_{\Lambda}
\right]dudv \quad .
\label{hatH}
\end{eqnarray}
As discussed in Section 4, the general solution to the VTD
evolution equations can be written out explicitly. In fact, for
the purpose of showing that the polarized and the half-polarized
$U(1)$ Symmetric solutions have AVTD behavior, we only need a
large $\tau$ expression for the VTD solution. For the polarized
case $(r = 0, w = 0)$, we have the following:

\subsubsection*{Large $\tau$ VTD Solution (Polarized)}

\begin{eqnarray}
\hat{\phi}(u,v,\tau) = \stackrel{\circ}{\phi} (u,v) -
v_{\phi}(u,v)\tau
\label{9a}
\end{eqnarray}
\begin{eqnarray}
\hat{p}(u,v,\tau) = -4v_{\phi}(u,v)
\label{9b}
\end{eqnarray}
\begin{eqnarray}
\hat{\Lambda}(u,v,\tau) = \stackrel{\circ}{\Lambda} (u,v) + 2
\tau - v_{\Lambda}(u,v)\tau
\label{9c}
\end{eqnarray}
\begin{eqnarray}
\hat{p}_{\Lambda}(u,v,\tau) = v_{\Lambda}(u,v)
\label{9d}
\end{eqnarray}
\begin{eqnarray}
\hat{x}(u,v,\tau) = \stackrel{\circ}{x}(u,v)
\label{9e}
\end{eqnarray}
\begin{eqnarray}
\hat{p}_x(u,v,\tau) = v_x(u,v)
\label{9f}
\end{eqnarray}
\begin{eqnarray}
\hat{z}(u,v,\tau) = \stackrel{\circ}{z}(u,v) - v_z(u,v)\tau
\label{9g}
\end{eqnarray}
\begin{eqnarray}
\hat{p}_z(u,v,\tau) = -4v_z(u,v)
\label{9h}
\end{eqnarray}
Here $v_{\phi}$ and $v_z$ are strictly positive \footnote {This
parametrization of the large $\tau$ VTD solution requires this
sign condition.}, and the notation $\hat{\phi},\hat{p}$, etc, is
used to indicate that these are VTD solutions, not Einstein
solutions. The eight functions $\{\stackrel{\circ}{\phi},
v_{\phi},
\stackrel{\circ}{\Lambda}, v_{\Lambda}, \stackrel{\circ}{x}, v_x,
\stackrel{\circ}{z}, v_z\}$ are all -- until we consider the VTD
constraints, in Section 4 -- treated as free functions on $T^2$;
they parameterize the set of solutions of the large $\tau$ VTD
equations corresponding to $\hat{H}$ in equation (\ref{hatH})

We now write the polarized $U(1)$ Symmetric fields each as the
sum of the appropriate term from the large $\tau$ VTD solution,
plus a remainder term. Specifically, for $\epsilon 0$, we have

\subsubsection*{Expansion Ansatz (Polarized)}
\begin{eqnarray}
\phi = \hat{\phi} + \delta \phi
\label{10a}
\end{eqnarray}
\begin{eqnarray}
p = \hat{p} + e^{-\epsilon \tau} \delta p
\label{10b}
\end{eqnarray}
\begin{eqnarray}
\Lambda = \hat{\Lambda} + \delta \Lambda
\label{10c}
\end{eqnarray}
\begin{eqnarray}
p_{\Lambda} =
\hat{p}_\Lambda + e^{-\epsilon \tau} \delta p_\Lambda
\label{10d}
\end{eqnarray}
\begin{eqnarray}
x = \hat{x}+ \delta x
\label{10e}
\end{eqnarray}
\begin{eqnarray}
p_x = \hat{p}_x + e^{-\epsilon \tau} \delta p_x
\label{10f}
\end{eqnarray}
\begin{eqnarray} z = \hat{z} + \delta z
\label{10g}
\end{eqnarray}
\begin{eqnarray}
p_z = \hat{p}_z + e^{-\epsilon \tau} \delta p_z
\label{10h}
\end{eqnarray}

The aim now is to substitute this ansatz into the Einstein
evolution equations, and show that the set of remainder functions
$\{\delta \phi, \delta p, \cdots, \delta p_z\}$, collectively,
satisfies an evolution equation in Fuchsian form. That is, if one
uses $\Psi$ to denote a vector whose components consist of
$\{\delta \phi, \delta p, \cdots, \delta p_z\}$ and certain of
their spatial derivatives, then one wants to show that $\Psi$
satisfies the Fuchsian equation\footnote{In terms of $t =
e^{-\tau}$, equation (\ref{11}) takes the more familiar Fuchsian
form
\begin{eqnarray*}
t\partial_t \Psi + A \Psi = -t^{\mu}F (t,
x, \Psi, \nabla\Psi).
\label{footnoteequation}
\end{eqnarray*}}
\cite{K96} \cite{KR98}
\begin{eqnarray}
\partial_{\tau} \Psi - A \Psi = e^{-\mu \tau} F (\tau, u,v,\Psi,
\nabla\Psi)
\label{11}
\end{eqnarray}
where $A$ is a matrix which is independent of $\tau$, and for
which \newline
$\sigma^A [:= \exp ( A\, \ell n \, \sigma)]$ is uniformly
bounded for $0 < \sigma < 1$; where $\mu$ is a strictly positive
constant; and where $F$ is continuous in $\tau$, is analytic in
$u,v$ and $\Psi$, and satisfies an estimate of the form
\begin{eqnarray}
\left| F (\tau, u,v,\Psi, \nabla\Psi) - F
(\tau, u,v, \Theta, \nabla\Theta)\right| \leq C \left[ \left|
\Psi - \Theta \right| + \left|\nabla\Psi
-\nabla\Theta\right|\right] \quad ,
\label{12}
\end{eqnarray}
for some constant C, provided that $\Theta$, $\nabla \Theta$,
$\Psi$, and $\nabla\Psi$ are bounded. ( Here $``| |"$ refers to
the pointwise absolute value norm, summed over the relevant
components of $F, \Psi,$ etc.)

For our present purposes, one readily verifies that it is
sufficient to show that each of the functions $\delta\psi, \delta
p$, etc (generically labeled ``$q$'') satisfies an evolution
equation of the form
\begin{eqnarray}
\partial _{\tau} q - \nu q
= \sum_{k} e^{- \mu_{k} \tau} f_k (\tau, u,v, \Psi, \nabla\Psi)
\label{201}
\end{eqnarray}
where the sum is finite; where $\nu(u,v)\geqslant 0$; where the
$\mu_k(u,v)$ are $\tau$-indepen\l-dent functions, each bounded
from below by a strictly positive constant; and where the
functions
$f_k$ are analytic in $u, v$, $\tau$, $\Psi$ and $\nabla\Psi$, and
are bounded by a polynomial in $\tau$.

We shall now show explicitly that the evolution equations for
$\delta \phi$ and $\delta p$ take this form, and argue that the
evolution equations for the other six functions $\delta \Lambda$,
etc, do as well.

We start with the evolution equation for $\delta \phi$. Based on
the Einstein Hamiltonian equation
\begin{eqnarray}
\partial _{\tau}\phi = {1 \over 4}p,
\label{13}
\end{eqnarray}
which we derive from the Hamiltonian (\ref{6a}) and (\ref{6b}),
we obtain
\begin{eqnarray}
\partial _{\tau}\delta\phi = {1 \over 4}e^{-\epsilon\tau} \delta p \quad,
\label{14}
\end{eqnarray}
where we recall that $\epsilon$ is the strictly positive
parameter appearing in the expansion ansatz
(\ref{10a})-(\ref{10h}). Clearly this is of the right form, with
$\nu = 0$, $\mu_1 = \epsilon$, and $f_1 = \delta p$.

The equation for $\delta p$ is not so simple. Based on the
Einstein Hamiltonian equation
\begin{eqnarray}
\partial _{\tau} p = 4e^{-2\tau}
\left(e^{\Lambda}e^{ab}\phi,_a \right) ,_b \label{15}
\end{eqnarray}
we derive
\begin{eqnarray}
\partial _{\tau}\delta p - \epsilon\delta p =
4\left[e^{\stackrel{\circ}{\Lambda} + \epsilon\tau - v_{\Lambda}\tau +
\delta\Lambda}
e^{ab}\left(\stackrel{\circ}{\phi},_a-v_{\phi,a}\tau + \delta \phi,_a
\right)
\right],_b
\label{16}
\end{eqnarray}
ith $e^{ab}$ the inverse of $e_{ab}$ from (\ref{5}); i.e.,
\begin{eqnarray}
e^{ab} = {1 \over 2}
\left(\begin{array}{cc}
e^{2z}+e^{-2z}(1-x)^2&-e^{2z}-e^{-2z}\left(x^2-1\right)\\
-e^{2z}-e^{-2z}\left(x^2-1\right)&
e^{2z}+e^{-2z}(1+x)^2
\end{array}\right) \quad ,
\label{17}
\end{eqnarray}
and with $x$ and $z$ to be expanded as in equations (\ref{10a})-(\ref{10h})
and (\ref{9a})-(\ref{9h}). Let us now write $e^{ab}$ in the form
\begin{eqnarray} e^{ab} &=& e^{-2z}{1 \over 2}\left(\begin{array}{cc}
\left(1-x\right)^2&-\left(x^2-1\right)\nonumber\\ -\left(x^2-1\right)&
(1+x)^2\end{array}\right) + e^{2z}{1 \over 2}\left(\begin{array}{cc}
1&-1\nonumber\\ -1&1\end{array}\right)\nonumber\\ &=:& e ^{-2z}e_{I}^{ab} +
e^{2z}e_{II}^{ab}
\label{18}
\end{eqnarray}
The right hand side of (\ref{16}) splits, and we have
\begin{eqnarray}
\partial _{\tau} p - \epsilon\delta p &=& 4
\left\{e^{\stackrel{\circ}{\Lambda} +
\epsilon\tau - v_{\Lambda}\tau + \delta\Lambda}
e^{-2\stackrel{\circ}{z} + 2v_z\tau - 2\delta z} e_I ^{ab}
\left[\stackrel{\circ}{\phi},_a-v_{\phi,a}\tau + \delta \phi,_a
\right] \right\},_b \nonumber\\ &&\mbox{}+ 4
\left\{e^{\stackrel{\circ}{\Lambda} + \epsilon\tau -
v_{\Lambda}\tau +
\delta\Lambda} e^{+2\stackrel{\circ}{z} - 2v_z\tau + 2\delta z}
e_{II} ^{ab}
\left[\stackrel{\circ}{\phi},_a-v_{\phi,a}\tau + \delta
\phi,_a\right]\right\} ,_b \quad \quad
\label{19}
\end{eqnarray}
Calculating out the first term on the right hand side of equation
(\ref{19}), we obtain
\begin{eqnarray}
RHS_I = e^{\left(\epsilon
- v_{\Lambda}+2v_z \right)\tau} 4e^{\stackrel{\circ}{\Lambda} -
2\stackrel{\circ}{z}}e^{\delta\Lambda - 2 \delta z}\left\{e_I
^{ab},_b \left(\stackrel{\circ}{\phi},_a-v_{\phi,a}\tau + \delta
\phi,_a\right)\right.\nonumber\\
+ e_I^{ab} \left(\stackrel{\circ}{\phi},_{ab}-v_{\phi,ab}\tau +
\delta \phi,_{ab}\right)\nonumber\\
+ \left[\stackrel{\circ}{\Lambda},_b -
2\stackrel{\circ}{z},_b -(v_{\Lambda},_b - 2v_{z,_b})\tau +
\delta \Lambda,_b - 2\delta z,_b
\right] \nonumber\\
\left.  e_I ^{ab}
\left(\stackrel{\circ}{\phi},_{a}-v_{\phi,a}\tau + \delta
\phi,_a\right) \right\}.
\label{20}
\end{eqnarray}
This expression contains second derivatives of some of the fields
included in $\Psi$, and so it cannot satisfy the conditions
demanded for $f_k$ from equation (\ref{201}) as listed above.
However, by introducing a new variable
\begin {eqnarray}
\phi_a \equiv \phi,_a
\label{phia}
\end {eqnarray}
together with its corresponding expansion with remainder term,
\begin {eqnarray}
\phi_a =
\stackrel{\circ}\phi,_a-v_{\phi,a}\tau+ \delta\phi_a,
\label{expphia}
\end{eqnarray}
and by including $\delta \phi_a$ in an expanded version of
$\Psi,$ one readily transforms (\ref{19}) into a function
involving only first derivatives of the expanded $\Psi,$ as
required.

Now, inspecting the terms in (\ref{20}), including those in the matrix
$e_I^{ab}$, we see that, so long as the functions $v_{\Lambda}$ and $v_z$
satisfy the condition \begin{eqnarray} v_{\Lambda} > 2 v_z + \epsilon
\label{21} \end{eqnarray} for $\epsilon > 0$ as introduced above (equations
(\ref{10a})-(\ref{10h}), then $RHS_I$ takes the form required for the right
hand side of (\ref{201}), with $\mu_I := v_{\Lambda}- 2 v_z - \epsilon$.
Calculating the second term on the right hand side of (\ref{19}), we get a
very similar expression, only now we have $\mu_{II} := v_{\Lambda} + 2 v_z -
\epsilon$ and the condition for $\mu_{II}$ to be strictly positive is
\begin{eqnarray} v_{\Lambda} > -2 v_z + \epsilon \quad . \label{22}
\end{eqnarray} Assuming that $v_z$ is strictly positive (which is a
requirement for this expansion ansatz), (\ref{21}) is the more rigorous
condition. Thus we find that in this first step of checking whether
solutions of the ansatz form (\ref{10a})-(\ref{10h}) are to be AVTD, it is
sufficient that condition (\ref{21}) hold.

We proceed to consider the rest of the evolution equations for
$\delta\Lambda$, $\delta p_{\Lambda}$, etc (including the added
first derivative variables). We get very similar results. For
example, the $\delta\Lambda$ evolution equation is
\begin{eqnarray}
\partial_{\tau}\delta\Lambda = -e^{-\epsilon \tau}\delta
p_{\Lambda},
\label{23}
\end{eqnarray}
while the $\delta p_{\Lambda}$ equation - derived from the
evolution equation for $p_{\Lambda}$,
\begin{eqnarray}
\partial_{\tau}p_{\Lambda} =&& - e^{-2\tau}\left\{e^{\Lambda}
e^{ab}\Lambda,_{ab}+ \left(e^{\Lambda} e^{ab}
\right),_{ab}\right.\nonumber\\ &&+ e^{\Lambda}
\left.\left[\left(e^{-2z}\right),_u x,_v -
\left(e^{-2z}\right),_vx,_u\right] + 2e^{\Lambda}e^{ab} \phi,_a
\phi,_b \right\}
\label{24}
\end{eqnarray} --can be written as
\begin{eqnarray}
\partial_{\tau}\delta p_{\Lambda} -\epsilon \delta p_{\Lambda}&=&-
e^{\left(\epsilon - v_{\Lambda} + 2v_z\right)\tau}
e^{\stackrel{\circ}{\Lambda} - 2\stackrel{\circ}{z}}e^{\delta\Lambda
-2\delta z}\left\{e_I^{ab}\left[\stackrel{\circ}{\Lambda},_{ab}-
v_{\Lambda},_{ab}\tau +\delta\Lambda,_{ab} \right]\right. \nonumber\\ &&+
e_I^{ab}\left(\stackrel{\circ}{\Lambda}, _a-v_{\Lambda,a}\tau
+\delta\Lambda,_a
\right)\left(\stackrel{\circ}{\Lambda}, _b-v_{\Lambda,b}\tau
+\delta\Lambda,_b \right)\nonumber\\
&&+2e_I^{ab}\left(\stackrel{\circ}{\phi}, _a-v_{\phi,a}\tau +\delta\phi,_a
\right)\left(\stackrel{\circ}{\phi}, _b-v_{\phi,b}\tau +\delta\phi,_b
\right)\nonumber\\ &&\left.\mbox{ + similar terms}\right\}
\hspace{1.75in}\nonumber\\ &&- e^{\left(\epsilon - v_{\Lambda} -
2v_z\right)\tau} e^{\stackrel{\circ}{\Lambda} +
2\stackrel{\circ}{z}}e^{\delta\Lambda + 2\delta z}
\left\{e_{II}^{ab}\left[\mbox{etc}\right]+ \cdots \right\}. \label{25}
\end{eqnarray} We thus find that for this pair, too, so long as the
inequality (\ref{21}) holds, and so long as we define new field variables
\begin{eqnarray} \Lambda_a\equiv \Lambda,_a ,
\label{lambdaa}
\end {eqnarray}
\begin{eqnarray}
x_a\equiv x,_a ,
\label {xa}
\end{eqnarray}
\begin{eqnarray}
z_a\equiv z,_a ,
\label {za}
\end {eqnarray}
with corresponding expansions \begin{eqnarray} \Lambda_a =
\stackrel o\Lambda,_a -v_\Lambda,_a \tau+\delta\Lambda_a,
\label{lambdaaexp} \end {eqnarray}
\begin{eqnarray}
x_a =
\stackrel o x,_a +\delta x_a,
\label{xaexp}
\end{eqnarray}
\begin{eqnarray}
z_a =
\stackrel o z,_a -v_z,_a \tau+\delta z_a,
\label{zaexp}
\end{eqnarray}
the evolution equations are of the
proper (Fuchsian) form. The same inequality condition
(\ref{21}), together with the condition
\begin{eqnarray}
v_z>\epsilon/4
\label {v-z}
\end {eqnarray}
leads to the proper form for the evolution equations for the
quantities $\delta z$, $\delta p_z$, $\delta x$, and $\delta p
_x$, as well as for $\delta \phi_a$, $\delta\Lambda_a$, $\delta
x_a$, and $\delta z_a$. The evolution equations for these latter
four quantities are derived >from their definitions. For
example, the evolution equation for
$\delta
\phi_a$ is
\begin{eqnarray}
\partial_\tau \delta\phi_a=\frac{1}{4}e^{-\epsilon\tau }
\partial_a\delta p.
\end{eqnarray}

We have now verified the following:\\

\begin{proposition}
If the polarized $U(1)$ Symmetric
gravitational variables
\begin{eqnarray*} {\cal G}_{pol} :=
\left\{\phi, \phi_a, p, \Lambda, \Lambda_a, p_{\Lambda}, x, x_a,
p_x, z, z_a, p_z \right\}
\label{proposition1b}
\end{eqnarray*}
are expanded as in equations (\ref{9a})-(\ref{9h}),
(\ref{10a})-(\ref{10h}) , (\ref{expphia}), and
(\ref{lambdaaexp})-(\ref{zaexp}), then for any choice of the
asymptotic data
\begin{eqnarray*}
{\cal A}_{pol} :=
\left\{\stackrel{\circ}{\phi}, \stackrel{\circ}{\Lambda},
\stackrel{\circ}{x}, \stackrel{\circ}{z}, v_{\phi}, v_{\Lambda},
v_x, v_z\right\}
\label{proposition1a}
\end{eqnarray*}
which satisfies the conditions
\begin{eqnarray} v_z > \epsilon/4
\label{29c}
\end{eqnarray} and
\begin{eqnarray} v_{\Lambda} > 2v_z + \epsilon ,
\label{29d}
\end{eqnarray}
(for $\epsilon >0$), together with the
expansion ansatz condition
\begin{eqnarray}
v_{\phi}>0 ,
\label{29e}
\end{eqnarray}
the vacuum Einstein evolution equations take the form of a
Fuchsian system for
\begin{eqnarray*}
\delta{\cal G}_{pol} :=
\left\{\delta\phi, \delta\phi_a, \delta p, \delta\Lambda, \delta
\Lambda_a, \delta p_{\Lambda}, \delta x, \delta x_a, \delta p_x,
\delta z, \delta z_a, \delta p_z\right\} \quad .
\label{proposition}
\end{eqnarray*}
\end{proposition} It then follows from the work of Kichenassamy
and Rendall \cite{KR98} that we have \begin{corollary} For each
analytic choice of the asymptotic data ${\cal A}_{pol}$ which
satisfies conditions (\ref{29c})-(\ref{29e}), there is a unique
analytic solution $ \delta{\cal G}_{pol}$ to the vacuum Einstein
evolution equations for sufficiently large
$\tau$, with $ \delta{\cal G}_{pol}$ approaching zero as $\tau$
approaches infinity.
\end{corollary}

Hence we have an ${\cal A}_{pol}$-parameterized family of
asymptotically velocity term dominated solutions to the $U(1)$
symmetric vacuum Einstein evolution equations.

We have so far ignored the Einstein constraint equations. We shall consider
them in Section 4. In the rest of this section, we show that there is
another asymptotic expansion ansatz which allows for nonzero $r$ and $w$
and yet still shows AVTD behavior. It includes the polarized ansatz as a
subcase; and it introduces one extra free function into the asymptotic
data, rather than two. Hence we call these ``half-polarized'' solutions.

The half-polarized solutions are defined by their asymptotic form, which
appends the following expansions to those of the polarized solutions given
in (\ref{9a})-(\ref{9h}) and (\ref{10a})-(\ref{10h}):

\subsubsection*{Expansion Ansatz ( Half-Polarized)}

\begin{eqnarray}
r(u, v, \tau) = \stackrel{\circ}{r}(u, v) + e^{-\epsilon \tau}
\delta r
\label{30a}
\end{eqnarray}
\begin{eqnarray}
w(u, v, \tau) = e^{4
\left(\stackrel{\circ}{\phi} - v_{\phi}\tau + \delta\phi
\right)}\left[-{\stackrel{\circ}{r} \over 4v_\phi} + \delta w
\right] \quad .
\label{30b}
\end{eqnarray}
These expansions introduce one new function,
$\stackrel{\circ}{r}(u,v)$, into the set ${\cal A}_{1/2}$ of
asymptotic data, and include two time-dependent expansion
functions $\delta w (u, v, \tau)$ and $\delta r (u, v, \tau)$.

To verify that Proposition 1 and its Corollary 1 extend to the
half-polarized ansatz, we first expand out the evolution
equations for $\delta w$ and $\delta r$. For $\delta w$, it
follows from the Einstein Hamiltonian evolution equation
\begin{eqnarray}
\partial_{\tau} w = e ^{4 \phi} r
\label{31}
\end{eqnarray}
that we have
\begin{eqnarray}
\partial_{\tau} \delta w - 4 v_{\phi} \delta w = e ^{-\epsilon \tau}
\left[\left({\stackrel{\circ}{r} \over 4v_\phi} - \delta w \right) \delta p
+
\delta r\right]\quad ,
\label{32}
\end{eqnarray}
which is appropriate Fuchsian form, so long as $v_\phi\geq0$ (a
necessary condition for the expansion ansatz). For $\delta r$,
it follows from the Einstein Hamiltonian evolution equation
\begin{eqnarray}
\partial_{\tau} r =
e^{-2\tau}\left(e^{\Lambda}e^{-4\phi}e^{ab}w,_a \right),
_b
\label{33}
\end{eqnarray}
that so long as condition
(\ref{29d}) is satisfied, and so long as we introduce first
derivative field variables for $w$, we have Fuchsian form as
well.

We still need to verify that the evolution equations for $\delta
\phi$, $\delta p$, and the other remaining functions in $\delta
{\cal G}_{1/2}$ retain Fuchsian form. The evolution equations
for $\delta \phi$, $\delta \Lambda$, $\delta x$, and $\delta z$,
as well as for their first derivative field variables $\delta
\phi_a$, $\delta \lambda_a$, $\delta x_a$ and $\delta z_a$ are
unchanged by the nonvanishing of the $r$ and $w$ terms in the
Hamiltonian (\ref{6a})-(\ref{6b}), so they are in proper form.
The evolution equations for $\delta p$, $\delta p_{\Lambda}$,
$\delta p_x$, and $\delta p_z$ all pick up extra terms which
need to be checked. For example, the equation for $\delta p$,
derived from the Einstein Hamiltonian evolution equation
\begin{eqnarray}
\partial _{\tau}p &=& 4e^{-2\tau}
\left(e^{\Lambda}e^{ab}\phi,_a \right),_b\nonumber\\
&-&2e^{4\phi} r^2 + 2e^{-2\tau}e^{\Lambda}e^{-4\phi}e^{ab}w,_a
w,_b
\label{34}
\end{eqnarray}
takes the form
\begin{eqnarray}
\partial _{\tau}\delta p - \epsilon \delta p = 4 \left[
e^{\stackrel{\circ}{\Lambda}+ \left(\epsilon -
v_{\Lambda}\right){\tau}+\delta
\Lambda}e^{ab}\left(\stackrel{\circ}{\phi},_a - v_{\phi ,a}\tau
+ \delta \phi ,_a
\right)\right], _b\nonumber\\ -2e^{4\stackrel{\circ}{\phi}+
\left(\epsilon - 4\stackrel{\circ}{v}_\phi \right)\tau + 4
\delta \phi}\left(\stackrel{\circ}{r} + e^{-\epsilon \tau}\delta
r \right)^2\nonumber\\ +2 e^{\stackrel{\circ}{\Lambda} +
\left(\epsilon - v_{\Lambda} \right)\tau + \delta {\Lambda}}
e^{-4 \stackrel{\circ}{\phi} + 4v_{\phi}\tau - 4\delta
\phi}e^{ab} \times\nonumber\\ \times
\left(e^{4\stackrel{\circ}{\phi} - 4v_{\phi}\tau + 4\delta \phi}
\left[{-\stackrel{\circ}{r} \over 4v_{\phi}} + \delta
w\right]\right), _a \nonumber\\
\times \left(e^{4\stackrel{\circ}{\phi} - 4v_{\phi}\tau +
4\delta \phi} \left[{-\stackrel{\circ}{r} \over 4v_{\phi}} +
\delta w\right]\right), _b . \label{35}
\end{eqnarray}

The first term on the right hand side of equation (\ref{35}) is
just the right hand side of equation (\ref{16}), the evolution
equation for $\delta p$ in the polarized case; it has already
been checked. The second term, a new one, clearly is in Fuchsian
form so long as
\begin{eqnarray} v_{\phi} > {1
\over 4} \epsilon \quad .
\label{36}
\end{eqnarray}
The third term can be written in the form
\begin{eqnarray}
2e^{\left(\epsilon - v_{\Lambda} + 2v_z - 4v_{\phi}\right)\tau}
e^{\stackrel{\circ}{\Lambda} - 2 \stackrel{\circ}{z} +
4\stackrel{\circ}{\phi}} e^{\delta \Lambda - 2\delta z + 4
\delta \phi} \times \nonumber\\ \times
\left\{e_I^{ab}\left[4\left(\stackrel{\circ}{\phi},_a -
v_{\phi,a}\tau +
\delta \phi,_a\right)\left({-\stackrel{\circ}{r}\over 4v_{\phi}}
+ \delta w\right) + \left({-\stackrel{\circ}{r},_a \over
4v_{\phi}} + {\stackrel {\circ}r \over 4v_{\phi}^2} v_{\phi,a} +
\delta w,_a\right)\right]\right. \times \nonumber\\
\times
\left.\left[4\left(\stackrel{\circ}{\phi},_b - v_{\phi,b}\tau +
\delta \phi,_b\right)\left({-\stackrel{\circ}{r}\over
4v_{\phi}}+ \delta w \right) +
\left({-\stackrel{\circ}{r},_{b} \over 4v_{\phi}} + {\stackrel
{\circ}r \over 4v_{\phi}^2} v_{\phi,b} + \delta
w,_b\right)\right]\right\}\nonumber\\ + e^{\left(\epsilon -
v_{\Lambda} - 2v_z -
4v_{\phi}\right)\tau}e^{\stackrel{\circ}{\Lambda} + 2
\stackrel{\circ}{z} + 4\stackrel{\circ}{\phi}} e^{\delta \Lambda
+ 2\delta z + 4 \delta \phi}
\times \nonumber\\
\times \left\{e_{II}^{ab}\left[\mbox{same terms}\right]_a
\left[\mbox{same terms}\right]_b\right\}
\label{37}
\end{eqnarray}

So we find that this term--and consequently the half-polarized
evolution equation for $\delta p$--takes Fuchsian form so long as
\begin{eqnarray}
v_{\Lambda} > 2v_z - 4v_{\phi} + \epsilon \quad,
\label{38}
\end{eqnarray}
which follows from conditions
(\ref{29c})-(\ref{29e}). Examining the extra terms in the
evolution equations for $\delta p_{\Lambda}$, $\delta p_x$, and
$\delta p_z$, we reach the same conclusion. The full system of
evolution equations (including the first derivative variables
introduced above) takes Fuchsian form, and we have the following:

\begin{proposition} If the $U(1)$ Symmetric gravitational
variables
\begin{eqnarray*} {\cal G}_{1/2} := \left\{\phi,
\phi_a, p, \Lambda, \Lambda _a, p _{\Lambda}, x, x_a, p_x, z,
z_a, p_z, w, w_a, r\right\}
\label{proposition2a}
\end{eqnarray*}
are expanded as in equations (\ref{9a})-(\ref{10h}), then for any
choice of the asymptotic data
\begin{eqnarray*}
{\cal A}_{1/2} :=
\left\{\stackrel{\circ}{\phi}, \stackrel{\circ}{\Lambda},
\stackrel{\circ}{x}, \stackrel{\circ}{z}, v_{\phi}, v_{\Lambda},
v_x, v_z,\stackrel{\circ}{r}\right\}
\label{proposition2b}
\end{eqnarray*}
which satisfies conditions
(\ref{29c})-(\ref{29e}), the vacuum Einstein evolution equations
take the form of a Fuchsian system for
\begin{eqnarray*}
\delta{\cal G}_{1/2} := \left\{\delta \phi, \delta \phi_a, \delta
p, \delta \Lambda,
\delta \Lambda_a,\delta p _{\Lambda}, \delta x, \delta x_a,
\delta p_x, \delta z,
\delta z_a, \delta p_z, \delta w, \delta w_a, \delta r\right\}
\label{proposition2c}
\end{eqnarray*}
\end{proposition}
\begin{corollary} For each analytic choice of the asymptotic data
${\cal A}_{1/2}$ which satisfies (\ref{29c})-(\ref{29e}), there
is a unique solution $\delta{\cal G}_{1/2}$ to the vacuum
Einstein evolution equations for large $\tau$, with $\delta{\cal
G}_{1/2}$ approaching zero as $\tau$ approaches infinity.
\end{corollary}

\section{The Constraint Equations}

The results of section 3 show that for the polarized as well as
for the half-polarized families of $U(1)$ Symmetric spacetimes,
there is a function space of asymptotic data such that for any
choice of data in that function space there is a unique
corresponding solution of the $U(1)$ Symmetric Einstein evolution
equations which is AVTD and asymptotically approaches that choice
of data. These results ignore the Einstein constraint equations;
hence the spacetimes they describe are generally not solutions of
the Einstein vacuum field equations. We now wish to show that for
both of these families, there is a set of constraint equations
{\it on the asymptotic data} such that if a set of asymptotic
data satisfies these asymptotic constraint equations, then the
corresponding solution of the evolution equations satisfies the
full constraint equations (and therefore satisfies the full
Einstein vacuum field equations). In addition, we wish to show
that, for each of the two families, the asymptotic constraint
equations admit a set of solutions of the appropriate number of
parameters. We may then conclude that we have a corresponding set
of AVTD solutions of the Einstein vacuum field equations in each
of the two cases.

We note that, in addition to the constraint equations
${\mathcal{H}}_0=0$, ${\mathcal{H}}_u=0$, and
${\mathcal{H}}_v=0$, which we treat here, there are new
constraints on the initial data of the form $\phi_a$ =
$\partial_a \phi$ which are necessitated by the introduction of
the new variables $\phi_a$, $ \lambda_a$, $x_a$, and $z_a$ , as
discussed above. These readily translate into constraints on the
asymptotic data, and are preserved by the evolution equations, so
we shall not treat them here any further.

We begin the analysis of the constraints ${\mathcal{H}}_0=0$,
${\mathcal{H}}_u=0$, and ${\mathcal{H}}_v=0$ by deriving the
corresponding asymptotic constraint equations. We do this for
each family by substituting the appropriate expansion ansatz for
the fields into the constraint equations, and then letting $\tau
\rightarrow \infty$, with the consequent vanishing of all
$\delta$(field) terms and all terms consisting of decaying
exponentials times fields bounded by polynomials in $\tau$. More
specifically, starting with the polarized spacetimes, we
substitute expressions (\ref{10a})-(\ref{10h}) into (\ref{7a}),
(\ref{7b}), and (\ref{7c}). Then, setting $\delta x$, $\delta z$,
$\delta \phi$, $\delta \Lambda$ and their derivatives to zero,
and setting $e^{-4v_z\tau}$ and like terms to zero as well, we
obtain the following\footnote{In calculating
$\stackrel{\circ}{\mathcal{H}}_u$ and
$\stackrel{\circ}{\mathcal{H}}_v$, we have used (\ref{70}) to
cancel certain terms and thereby simplify the expressions
(\ref{71}) and (\ref{72})}.

\subsubsection*{Asymptotic Constraint Equations}

\begin{eqnarray}
0 &=& \stackrel{\circ}{\mathcal{H}}_0 \nonumber \\ &:=& 2
\left[\left(v_z\right)^2 + \left(v_{\phi}\right)^2 - {1 \over
4}\left(v_{\Lambda} \right)^2\right]
\label{70}
\end{eqnarray}
\begin{eqnarray}
0 &=& \stackrel{\circ}{\mathcal{H}}_u \nonumber \\
&:=&-4v_z \stackrel{\circ}{z},_u + v_x \stackrel{\circ}{x},_u +
v_{\Lambda} \stackrel{\circ}{\Lambda},_u - 4
v_{\phi} \stackrel{\circ}{\phi},_u - v_{\Lambda,u} \nonumber\\
& &\mbox{} + {1 \over 2} \left\{-\left(1 + \stackrel{\circ}{x}
\right)^2 v_x + 4 \left(1 + \stackrel{\circ}{x} \right) v_z
\right\},_v
\nonumber \\
& &\mbox{}-{1 \over 2}\left\{+\left(1 - \stackrel{\circ}{ x}^2
\right) v_x + 4 \stackrel{\circ}{x} v_z \right\},_u
\label{71}
\end{eqnarray}
\begin{eqnarray}
0 &=& \stackrel{\circ}{\mathcal{H}}_v \nonumber \\
&:=&-4v_z \stackrel{\circ}{z},_v + v_x \stackrel{\circ}{x},_v +
v_{\Lambda} \stackrel{\circ}{\Lambda},_v - 4
v_{\phi} \stackrel{\circ}{\phi},_v - v_{\Lambda,v} \nonumber\\
&&\mbox{}+{1 \over 2}\left\{\left(1 -
\stackrel{\circ}{x}\right)^2 v_x + 4
\left(1 - \stackrel{\circ}{x} \right) v_z \right\},_u \nonumber
\\ &&\mbox{}+{1
\over 2} \left\{\left(1 - \stackrel{\circ}{x}^2\right) v_x + 4
\stackrel{\circ}{x} v_z \right\},_v
\label{72}
\end{eqnarray}
These are constraints on the choice of the asymptotic data
${\mathcal A}_{pol}$ for polarized $ U(1)$ Symmetric spacetimes.
A similar procedure leads to constraints on ${\mathcal A}_{1/2}$,
the family of half-polarized
$U(1)$ Symmetric spacetimes. One finds that the asymptotic
constraints here are exactly (\ref{70}) -(\ref{72}); the terms
${1 \over 2} e^{4\phi}r^2$ and ${1 \over 2}
e^{\Lambda}e^{-4\phi}e^{ab}w,_a w,_b$ in ${\mathcal H}$ and the
terms $rw,_u$ in ${\mathcal H}_u$ and $rw,_v$ in ${\mathcal H}_v$
have no asymptotic effect if one expands $r$ and $w$ as in
(\ref{30a}) and (\ref{30b}).

We now want to argue that if the asymptotic data ${\mathcal
A}_{pol}$, or ${\mathcal A}_{1/2}$ \ satisfy the asymptotic
constraints (\ref{70})-(\ref{72}) and if the corresponding fields
${\mathcal G}_{pol}$ or ${\mathcal G}_{1/2}$ \ satisfy the $
U(1)$ Symmetric evolution equations generated by (\ref{6a}), then
${\mathcal G}_{pol}$, and ${\mathcal G}_{1/2}$ \ satisfy the
initial value constraints (\ref{7a})-(\ref{7c}) on any Cauchy
surface sufficiently close to the singularity (i.e., for
sufficiently large $\tau$). We argue this as follows (focusing
first on the polarized case): If we substitute the expansion
ansatz (\ref{10a})-(\ref{10h}) for the polarized fields into the
expressions (\ref{7a})-(\ref{7c}) for the constraints, then we
can write
\begin{eqnarray}
{\mathcal H}_0 =
\stackrel{\circ}{\mathcal H}_0 + e^{-\mu \tau} \delta {\mathcal
H}_0
\label{73}
\end{eqnarray}
\begin{eqnarray}
{\mathcal H}_u = \stackrel{\circ}{\mathcal H}_u + \delta
{\mathcal H}_u
\label{74}
\end{eqnarray}
\begin{eqnarray}
{\mathcal H}_v = \stackrel{\circ}{\mathcal H}_v + \delta
{\mathcal H}_v
\label{75}
\end{eqnarray}
where $\left\{{\mathcal H}_0, {\mathcal H}_u, {\mathcal
H}_v\right\}$ are the asymptotic constraint expressions
(\ref{70})-(\ref{72}), $\mu$ is a strictly positive constant, and
$\left\{\delta {\mathcal H}_0, \delta {\mathcal H}_u, \delta
{\mathcal H}_v\right\}$ are the remainder terms (defined by
(\ref{73})-(\ref{75})). Our calculations leading to
(\ref{70})-(\ref{72}) show that, regardless of whether the
asymptotic data ${\mathcal A}_{pol}$ satisfy the constraints
(\ref{70})-(\ref{72}), if the fields ${\mathcal G}_{pol}$ evolve
via the $U(1)$ Symmetric evolution equations, then the remainder
terms approach zero as $\tau \rightarrow \infty$. We want to show
more. We will show that in fact, so long as the asymptotic
constraints (\ref{70})-(\ref{72}) are satisfied, ${\mathcal
H}_0$, ${\mathcal H}_u$, and ${\mathcal H}_v$ are identically
zero for sufficiently large $\tau$. We do this by using the
$U(1)$ Symmetric evolution equations to show that
$\left\{{\mathcal H}_0, {\mathcal H}_u, {\mathcal H}_v\right\}$
are identically zero for sufficiently large $\tau$. To do that,
we use the U(1) Symmetric evolution equations to show that
$\delta {\mathcal H}_0, \delta {\mathcal H}_u, \delta {\mathcal
H}_v$ satisfy a Fuchsian system, and then note that this Fuchsian
system admits $\delta {\mathcal H}_0 = 0$, $\delta {\mathcal H}_u
= 0$, $\delta {\mathcal H}_v = 0$ as a solution. Thus, since the
solution to the Fuchsian system (for a given set of asymptotic
data) is unique near
$\tau \rightarrow \infty$ , and since $\left\{0, 0, 0\right\}$
is a solution, it follows that we have $\delta {\mathcal H}_0 =
0$, $\delta {\mathcal H}_u = 0$, $\delta {\mathcal H}_v = 0$ for
sufficiently large $\tau$. To complete the argument, we note
that if the asymptotic data ${\mathcal A}_{pol}$ satisfy the
asymptotic constraints and if
$\left\{\delta {\mathcal H}_0, \delta {\mathcal H}_u, \delta
{\mathcal H}_v \right\} = \left\{0, 0, 0\right\}$ for large
$\tau$, then it follows >from (\ref{73})-(\ref{75}) that
${\mathcal H}_0 = 0$, ${\mathcal H}_u = 0$, ${\mathcal H}_v = 0$
for large $\tau$. The constraints are satisfied.

The key to this argument is the derivation of the Fuchsian system
for $\left\{\delta {\mathcal H}_0, \delta {\mathcal H}_u,\delta
{\mathcal H}_v \right\}$. To obtain this system, we start with
the evolution equations for $\left\{{\mathcal H}_0, {\mathcal
H}_u, {\mathcal H}_v \right\}$, which can be calculated from the
divergence Bianchi identity:\footnote{Note that while ${\mathcal
H}_0$ is a weight two scalar density, ${\mathcal H}_u$,
${\mathcal H}_v$ and ${\mathcal H}_3$ are all weight one scalar
densities. Note also that while we have already eliminated
${\mathcal H}_3 = 0$ (see equation (\ref{3})), we leave
${\mathcal H}_3$ in equations (\ref{77})-(\ref{79}) for the sake
of generality.} \begin{eqnarray} \partial_{\tau} {\mathcal H}_0
= -e^{\Lambda-2\tau}[\Lambda,_b e^{ab}({\mathcal
H}_a-\beta_a{\mathcal H}_3) +(e^{ab}({\mathcal H}_a-
\beta_a{\mathcal H}_3)),_b]
\label{Hzero}
\end {eqnarray}
\begin{eqnarray}
\partial_{\tau} {\mathcal H}_a = - \partial_a {\mathcal H}_0
\label{77}
\end{eqnarray}
where $a, b$ run over $u$ and $v$. If we now substitute the
expansions (\ref{73})-(\ref{75}) into these evolution equations,
and further set
$\stackrel{\circ}{\mathcal H}_0, \stackrel{\circ}{\mathcal H}_u,
\stackrel{\circ}{\mathcal H}_v$ and ${\mathcal H}_3$ equal to
zero, then we derive
\begin{eqnarray}
\partial_{\tau} \delta {\mathcal H}_0 - \mu \delta {\mathcal
H}_0 = \left(e^{\Lambda-2\tau+\mu
\tau}\right)\left[\Lambda,_b e^{ab} \delta {\mathcal H}_a +
\left(e^{ab} \delta {\mathcal H}_a \right),_b\right] \label{78}
\end{eqnarray}
\begin{eqnarray}
\partial_{\tau} \delta {\mathcal H}_a = -e^{-\mu {\tau}}
\partial_a \left(\delta{\mathcal H}_0\right)
\label{79}
\end{eqnarray}
These equations clearly constitute a Fuchsian
system for $\left\{\delta {\mathcal H}_0, \delta {\mathcal
H}_u,\delta {\mathcal H}_v \right\}$ so long as $\mu > 0$ and
$v_{\Lambda}> 2v_z + \mu$, which are the familiar inequality
conditions on the asymptotic data. Further, we note that
$\left\{\delta {\mathcal H}_0, \delta {\mathcal H}_u, \delta
{\mathcal H}_v \right\} = \left\{0, 0, 0\right\}$ is a solution
of the system (\ref{78})-(\ref{79}). Finally, we note that
(\ref{78})-(\ref{79}) holds not just for the polarized $ U(1)$
Symmetric spacetimes, but in fact for the half-polarized $ U(1)$
Symmetric spacetimes as well. Thus we have proven the following:

\begin{proposition}
If the asymptotic data for either a polarized
or a half-po\-larized U(1) Symmetric spacetime is analytic and
satisfies the asymptotic constraint equations
(\ref{70})-(\ref{72}), then the corresponding spacetime satisfies
the full set of Einstein constraint equations--and hence the full
set of Einstein vacuum field equations--for sufficiently large
$\tau$.
\end{proposition}

While Proposition 3 tells us that any solution of the asymptotic
constraint equations (\ref{70})-(\ref{72}) leads to a
solution\footnote{In fact, since $r_0$ and $w_0$ do not appear in
the asymptotic constraints, every solution of
(\ref{70})-(\ref{72}) leads to a large set of half-polarized
solutions, along with one polarized solution.} of the vacuum
field equations, it does not tell us anything about finding and
parameterizing solutions of the asymptotic constraints. We could
work directly with (\ref{70})-(\ref{72}); however, since they are
specified on a possibly degenerate manifold, we choose an
alternative approach.

The alternative approach starts with the recognition that
(\ref{70})-(\ref{72}) are the asymptotic constraint equations for
the $U(1)$ Symmetric VTD equations as well as for the $U(1)$
Symmetric vacuum Einstein equations. To see this, we substitute
the large $\tau$ VTD solutions--(\ref{9a})-(\ref{9h}) for the
polarized case, and (\ref{9a})-(\ref{9h}) together with $r =
\stackrel{\circ}{r}$ and $w = 0$ for the half-polarized
case--into the VTD constraint equations:
\begin{eqnarray}
0&=&\hat{\mathcal H}_0 \nonumber\\ &=&{1\over8} p^2_z +
{1\over2}e^{4z}p^2_x+{1\over8} p^2 + {e^{4\phi}\over2}r^2 -
{1\over2}p^2_{\Lambda}
\label{80}
\end{eqnarray}
\begin{eqnarray}
0&=&\hat{\mathcal H}_u \nonumber\\ &=&{\mathcal H}_u \quad
\mbox{(see equation (\ref{7b}))}
\label{81}
\end{eqnarray}
\begin{eqnarray}
0&=&\hat{\mathcal H}_v \nonumber\\ &=&{\mathcal
H}_v \quad \mbox{(see equation (\ref{7c}))}
\label{82}
\end{eqnarray}
We obtain, in both cases, equations
(\ref{70})-(\ref{72}).

We next note that we have global existence in time $\tau$ for
solutions of the $U(1)$ Symmetric VTD evolution equations. This
property holds even for the slightly generalized form of these
equations, which we obtain >from the Hamiltonian
\begin{eqnarray}
\hat{H} = \int_{T^2} \left[\alpha \hat{\mathcal H}_0+ 2
p_{\Lambda} \right]du dv ,
\label{83}
\end{eqnarray}
where $\alpha$ is a freely specifiable ``lapse'' function. Global
existence is an immediate consequence of the following expression
for the general solution to these evolution equations (clearly
well behaved for all values of $\tau \epsilon (-\infty, +
\infty))$ \cite{BM93} \cite{BM98a}:
\begin{eqnarray} \phi =
-v_{\phi} \left(\alpha \tau - \tau_{ \phi} \right) - {1 \over 2}
\ell n \left[\left|\zeta_{\phi} \right|\left(1 +
e^{-4v_{\phi}\left(\alpha\tau-\tau_{\phi} \right)}\right)\right]
\label{84}
\end{eqnarray}
\begin{eqnarray}
p = -4 v_{\phi} \left({1
-e^{-4v_{\phi}\left(\alpha\tau-\tau_{\phi} \right)}\over 1 +
e^{-4v_{\phi}\left(\alpha\tau-\tau_{\phi} \right)}}\right)
\label{85}
\end{eqnarray}
\begin{eqnarray}
w = \xi_{\phi}- \left[\zeta_{\phi} \left(1 +
e^{-4v_{\phi}\left(\alpha\tau-\tau_{\phi} \right)}
\right)\right]^{-1}
\label{86}
\end{eqnarray}
\begin{eqnarray}
r = -4\zeta_{\phi}v_{\phi}
\label{87}
\end{eqnarray}
\begin{eqnarray}
x = \xi_z - \left[ \zeta_z \left(1 +
e^{-4v_z\left(\alpha\tau-\tau_z \right)}\right) \right]^{-1}
\label{88}
\end{eqnarray}
\begin{eqnarray}
p_x = -4 \zeta_z v_z
\label{89}
\end{eqnarray}
\begin{eqnarray}
z = -v_z \left(\alpha\tau-\tau_z \right) - {1\over 2} \ell n
\left[\left|\zeta_{z}\right| \left(1 +
e^{-4v_z\left(\alpha\tau-\tau_z \right)}\right)\right]
\label{90}
\end{eqnarray}
\begin{eqnarray}
p_z = -4v_z \left( {1 - e^{-4v_z\left(\alpha\tau-\tau_z \right)}
\over 1 + e^{-4v_z\left(\alpha\tau-\tau_z\right)}}\right)
\label{91}
\end{eqnarray}
\begin{eqnarray}
\Lambda = \Lambda_0 + 2\tau -\alpha v_{\Lambda} \tau
\label{92}
\end{eqnarray}
\begin{eqnarray}
p_{\Lambda} = v_{\Lambda}
\label{93}
\end{eqnarray}
Here
$\tau_{\phi}$, $v_{\phi}>0$, $\zeta_{\phi}$, $\xi_{\phi}$,
$\tau_z$, $v_z>0$, $\zeta_z$, $\xi_{\phi}$, $\Lambda_0$ and
$v_{\Lambda}$ are all free functions of $u$ and $v$; they are
``constants of integration'' for the set of {\it ordinary}
differential equations which comprise the $ U(1)$ Symmetric VTD
evolution equation system. Note that (\ref{84})-(\ref{93}) is the
general solution for the {\it full} $ U(1)$ symmetric VTD
evolution system; no polarization or half-polarization assumption
has been made. Note also that the positivity conditions on
$v_{\phi}$ and $v_z$ are necessary for this parametrization of
the solutions of the VTD equations.

With global existence of the VTD solutions established, we may
consider solutions of the VTD constraints (\ref{80})-(\ref{82})
at finite times $\tau$, at which there is no difficulty with
degeneracy of the spatial geometry. To relate solutions of
(\ref{80})-(\ref{82}) to solutions of the asymptotic VTD
constraints (\ref{70})-(\ref{72}) (the same as the asymptotic
Einstein constraints), it is useful to establish the following
\begin{proposition} The VTD constraint functions $\hat{\mathcal
H}_0$, $\hat{\mathcal H}_u$, and $\hat{\mathcal H}_v$, when
evaluated for solutions (\ref{84})-(\ref{93}) of the VTD
evolution equations, are independent of $\tau$, so long as for
some $\tau_0$, $\hat{\mathcal H}_0(\tau_0) = 0$
\end{proposition}
$\mathbf {Proof}$: The time independence of the quantity
$\hat{\mathcal H}_0$ may be established directly, by substituting
expressions (\ref{84})-(\ref{93}) for the fields $\{\phi(\tau)$,
$p(\tau)$, $w(\tau)$,
$r(\tau)$, $x(\tau)$, $p_x(\tau)$, $z(\tau)$, $p_z(\tau)$,
$\Lambda(\tau), p_\Lambda (\tau) \}$, into the expression
(\ref{80}) for $\hat{\mathcal H}_0$. One obtains, for all $\tau$
(and all $\alpha$), $2v^2_z + 2v^2_{\phi} - {1 \over
2}v^2_{\Lambda}$, which is independent of $\tau$ (and equal to
$\hat{\mathcal H}_0$).

Rather than proceeding to substitute (\ref{84})-(\ref{93}) into
(\ref{81}) and (\ref{82}), we may establish the $\tau$
independence of $\hat{\mathcal H}_u$ and $\hat{\mathcal H}_v$
using the following argument: Since $\hat{\mathcal H}_u =
{\mathcal H}_u$ and $\hat{\mathcal H}_v = {\mathcal H}_v$, we
find that $\hat{\mathcal H}_u$ and $\hat{\mathcal H}_v$ both
generate spatial diffeomorphisms, tangent to $\tau =$ constant
surfaces. With $\hat{\mathcal H}_0$ constant and presumed zero at
time $\tau_0$, we have $\hat{\mathcal H}_0 = 0$ for all time.
Hence for any vector field $M^a(x)$ on $T^2$, we have
\begin{eqnarray}
\left\{\int_{T^2} M^a(x) \hat{\mathcal H}_a(x) dx,\int_{T^2}
\alpha(y) \hat{\mathcal H}_0(y) dy \right\} = 0
\label{94}
\end{eqnarray}
where $\left\{ \ , \right\}$ indicates the Poisson
bracket.

The generalized VTD Hamiltonian $\hat{H} $, from equation
(\ref{83}), generates VTD time evolution for $\hat{\mathcal H}_u$
and $\hat{\mathcal H}_v$ (as well as for any other function of
the VTD fields). As argued above, $\hat{\mathcal H}_u$ and
$\hat{\mathcal H}_v$ commute with $\int \alpha \hat{\mathcal
H}_0$. Calculating the Poisson bracket of $\hat{\mathcal H}_u$
and $\hat{\mathcal H}_v$ (times an arbitrary $M^a$, integrated
over $T^2$), with the remaining piece of the Hamiltonian
$\hat{H}$, we have
\begin{eqnarray}
\left\{ \int_{T^2} M^a(x) \hat{\mathcal H}_a(x) dx\right.,&&\left. 2
\int_{T^2} p_{\Lambda} (y) dy \right\} \nonumber \\ &=& \left\{-
\int_{T^2}\left(M^a p_{\Lambda} \right) , _a \Lambda dx, 2 \int_{T^2}
p_{\Lambda} (y) dy\right\} \nonumber\\ &=& -2 \int_{T^2}\left(M^a
p_{\Lambda} \right), _a dx
\nonumber\\ &=& 0
\label{95}
\end{eqnarray}
Thus $\hat{\mathcal H}_u$ and $\hat{\mathcal H}_v$ commute with
the VTD Hamiltonian, from which it follows that $\hat{\mathcal
H}_u$ and $\hat{\mathcal H}_v$ are VTD constant.

Our remaining task now is to show that one can find a full set of
solutions of the VTD constraint equations (\ref{80})-(\ref{82})
-- five\footnote{Three of these functions are related to
coordinate choice, and thus are not physical.} free functions on
$T^2$ in the polarized case; and six free functions as $T^2$ in
the half-polarized case. To do this, it is useful to first show
the following: For any fixed choice of the asymptotic data
${\mathcal A}_{pol}$ (or ${\mathcal A}_{1/2}$) there is a choice
of the lapse function $\alpha$ for which the VTD fields at $\tau =
1$ have constant mean curvature (CMC) in a 2 + 1 dimensional
sense; i.e., based on the 2 + 1 Lorentz signature metric
\begin{eqnarray}
\gamma = -\alpha^2e^{2\Lambda - 4 \tau} d\tau^2
+ e^{-2\tau + \Lambda} e_{ab} dx^a dx^b.
\label{96}
\end{eqnarray}
(compare with the 3+1 metric (1).) The spatial
volume element for $\gamma$ from (\ref{96}) is
\begin{eqnarray}
^2\mu = e^{\Lambda - 2\tau}
\label{97}
\end{eqnarray}
Using the
VTD evolution equations and solutions to calculate the mean
extrinsic curvature for a $\tau = $ constant surface in this 2 +
1 dimensional spacetime, we obtain
\begin{eqnarray}
tr K &=& {-1 \over (\mbox {lapse})} {\partial
_{\tau}\left(^2\mu\right) \over ^2\mu}\nonumber \\ &=& {v_\Lambda
\over e^{\Lambda - 2\tau}}\nonumber \\ &=& {v_\Lambda \over
e^{\Lambda_0 - \alpha\tau v_\Lambda}} .
\label{98}
\end{eqnarray}
If we now set $\tau = 1$ and $tr K = C>0$ in (\ref{98}), we can
solve for $\alpha$:
\begin{eqnarray}
\alpha = {1 \over v_{\Lambda}} \left(\Lambda_0 (x) - \ell n (\,
v_{\Lambda}(x)/C \right))
\label{99}
\end{eqnarray}
Noting that our main results concerning the
evolution behavior of the fields (Corollaries 1,2, and 3) hold
only if $v_{\Lambda}$ is positive definite, we see that indeed,
for any fixed data ${\mathcal A}_{pol}$ or ${\mathcal A}_{r}$, we
can choose $\alpha$ (as in (\ref{99})) so that the $\tau = 1$
hypersurface has constant mean curvature. We ensure the
positivity of $\alpha$ by a sufficiently large choice of the
constant $C$.

Now that we know that any VTD solution admits a hypersurface with
constant mean curvature in the 2+1 sense described above, we can
use this CMC condition to help in the analysis of the VTD
constraint equations, and show that we have a full set of
solutions. We carry out this analysis using the conformal method,
adapted to $ U(1)$ Symmetric data in 2+1 form, as in \cite{M86}
(where it is applied to the Einstein constraints). Specifically,
on
$T^2$, the manifold transverse to the U(1) orbits, we choose for
our conformal data (1) a Riemannian metric $\lambda_{ab}$, (2) a
symmetric tensor $\sigma^{cd}$ which is divergence-free and
trace-free with respect to $\lambda_{ab}$, (3) a constant $\tau$
representing the mean extrinsic curvature of the geometry on
$T^2$, and (4) a pair of functions $\tilde {\phi}$ and $\tilde
{p_{\phi}}$ which are treated as matter fields, but actually
correspond to the geometry in the direction tangent to the orbits
of the $U(1)$ isometry group. We then attempt to solve a system
of three equations for a vector field $W$ (which generates the
longitudinal part of the extrinsic curvature on $T^2$) and a
positive definite conformal factor $\psi$. If indeed for the
given set of conformal data $\{\lambda, \sigma, \tau,
\tilde{\phi}, \tilde{p}_{\phi} \}$, solutions $W$ and $\psi$
exist, then we may construct from $W$ and $\psi$ and the
conformal data a set of initial data
$ \{\gamma, \pi, \phi, p_\phi \} $, which satisfies the
constraints.

The equations to be solved for $W$ and $\psi$ are generally
coupled, and it is generally not easy to determine if solutions
exist. For CMC conformal data, however, the equations decouple;
and in the case of the Einstein constraints, it has been
determined that solutions to the equations exist for essentially
all choices of the conformal data. \cite{M86}

For the VTD constraints, of interest here, this decoupling still
holds. While there has not been any systematic study of the
equations for $W$ and $\psi$ resulting from the VTD constraints,
such a study is easily carried out. Indeed, the supermomentum
constraints are the same for the Einstein and the VTD cases; in
both cases, we have a linear elliptic system to be solved for
$W$, and in both cases a solution exists for all choices of
conformal data satisfying the integrability condition
\begin{eqnarray}
\int \left(\tilde{p} Y^a \nabla_a \tilde{\phi}\right) = 0
\end{eqnarray}
where $Y$ is any conformal Killing field on $T^2$. The
superHamiltonian constraint, to be solved for $\psi$, is
different for the Einstein and the VTD cases. In the former case,
it is a nonlinear elliptic equation for $\psi$, while in the
latter case, it is an algebraic equation for $\psi$. In both
cases, one verifies that for nonzero constant mean curvature and
for generic choices of the conformal data, a unique solution
exists. We note that if the conformal data is chosen to be
analytic, it follows from the ellipticity of the constraint
equations that the corresponding initial data is analytic. It
further follows from the VTD evolution equations and their
general solution that the corresponding asymptotic data is
analytic.

The analysis which we have just sketched shows that we have a
full set of analytic solutions of the VTD constraint equations,
and hence have a full set of analytic solutions of the asymptotic
Einstein constraints. If such solutions are to correspond to AVTD
solutions of the Einstein equations, we must verify that the
asymptotic data satisfies the inequalities $v_z>\epsilon/4$,
$v_{\phi}>\epsilon/4$, and $v_{\Lambda}>2 v_z+ \epsilon$, which
are sufficient for the evolution equations to be Fuschsian.

General solutions of the asymptotic constraints do ${\it not}$
satisfy these inequalities. We claim, however, that there are
open sets of choices of the conformal data which do guarantee
that these inequalities hold. To argue this, we first note that
as a consequence of the asymptotic constraint equation (\ref{70})
${v_{\Lambda}}^2=4({v_z}^2+{v_{\phi}}^2)$, of the compactness of
$T^2$, and of our freedom to choose $\epsilon$ to be any positive
constant, all of the necessary inequalities
(\ref{29c})-(\ref{29e}) hold (for some $\epsilon>0$) so long as
$v_{\phi}$ and $v_z$ are both positive definite.

Now, in finding data which satisfy the VTD constraints at finite
time ($\tau=1$), we have no apparent direct control over the
values of the data for arbitrarily large $\tau$ (ie, the values
of the asymptotic data). However, examining the explicit form of
the global-in-time VTD solutions (\ref{84}-\ref{93}), we find
that with $\tau=1$ and $\tau_{\phi}=0=\tau_z$, we have
\begin{eqnarray}
p=-4v_{\phi} \tanh(2\alpha v_{\phi}) \label{202}
\end{eqnarray}
and
\begin{eqnarray}
p_z=-4v_z \tanh(2 \alpha v_z) \label {203}
\end{eqnarray}
It follows from (\ref {202})-(\ref{203}) that for any set of
data $\{\phi, \Lambda, r, x, z, p, p_{\Lambda}, \omega, p_x,
p_z\}$ at $\tau=1$ which has $p<0$ and $p_z<0$ for all $u,v$ in
$T^2$, there exists a unique choice of positive definite
functions $v_{\phi}$ and $v_z$ such that (\ref {29c})-(\ref{29e})
hold. (The same is true for nonzero, but sufficiently small,
values of $\tau_\phi$ and $\tau_z$.) Thus each solution of the
VTD constraints at $\tau=1$ which has $p<0$ and $p_z<0$ has
corresponding asymptotic data which satisfies the inequalities
(\ref {29c})-(\ref{29e}).

The conformal procedure for solving the VTD constraints,
sketch\-ed above, allows us to freely choose the function $p$ at
$\tau=1$, and hence its sign; however, since the quantity $p_z$
is part of the extrinsic curvature of the two dimensional
geometry in our
$2+1$ treatment of the constraints, we cannot directly choose
$p_z$. Yet we can argue as follows that we can control the sign of
$p_z$: As just noted, $p_z$ is related to the extrinsic curvature
of the geometry on $T^2$. More explicitly, letting $\pi^{ab}$
denote the momentum representation of this extrinsic curvature,
we have
\begin{eqnarray}
p_z = [ \pi^{uu}(e^{2z}-e^{-2z} (1+x)^2)
+ \pi^{vv}(e^{2z}-e^{-2z}(1-x)^2)+ \nonumber\\
2\pi^{uv}(e^{2z}-e^{-2z}(x^2-1))]e^{\Lambda-2\tau}
\label{2000}
\end{eqnarray}
Now in solving the momentum constraints via the
conformal method, one determines the vector field $W$, which
partially determines $\pi^{ab}$. However, one can always add to
this part of $\pi^{ab}$ an arbitrary divergence-free, trace-free,
symmetric tensor density $\sigma^{ab}$. Since all metrics on
$T^2$ are conformally related to a flat metric, one finds that in
appropriate coordinates, $\sigma^{ab}$ is divergence-free and
trace-free if and only if $\sigma^a_b$ is a spatially constant
trace-free tensor density. It follows from this fact, and from
equation (99), that through the choice of $\sigma^{ab}$, one can
always guarantee that $p_z$ is negative definite. We then obtain,
as argued above, the positive definiteness of $v_z$

We conclude from this argument that, while it is not true that
all sets of VTD data satisfying the VTD constraints at time
$\tau=1$ lead to asymptotic data satisfying the inequalities, if
we impose certain open conditions on the choice of the conformal
data, then we do obtain VTD data whose asymptotic data satisfies
the inequalities. Noting the invertibility of functions such as
$f(v)=-4\tanh(2\alpha v)$ for $f<0$ and $\alpha v>0$, we readily
verify that for both the polarized and the half-polarized $U(1)$
Symmetric spacetimes, we have a full set of asymptotic data--two
free functions for the polarized solutions and three free
functions for the half polarized solutions (after quotienting out
the diffeomorphism gauge freedom on $T^2$)--which satisfy the
asymptotic constraints as well as the inequalities which we have
found to be sufficient to guarantee that the evolution equations
are Fuchsian. Combining this result with those of Section 3 we
have \begin{theorem} There is a family of U(1) Symmetric
solutions of the vacuum Einstein equations on $ T^3 \times R$
which is AVTD with respect to a harmonic time foliation, is
characterized by analytic asymptotic data ${\mathcal A}_{1/2}$
satisfying the asymptotic constraints (\ref{70})-(\ref{72}), and
is parametized by three free functions on $T^2$.

Those members of this family of spacetimes which have asymptotic
data with $\stackrel{\circ}{r}=0$ are AVTD solutions of the
polarized U(1) Symmetric vacuum Einstein equations. This
polarized subfamily is parametrized by two free functions on
$T^2$. \end{theorem}

\section{AVTD Behavior and the Choice of Observ\-ers}

The definition of AVTD behavior which we use here depends on a
choice of observers; or equivalently, on a choice of coordinates.
One may therefore ask if alternative sets of observers will agree
on the presence or absence of AVTD behavior in a given spacetime.
This issue has not yet been carefuly addressed, since in most
previous studies \cite{KR98} \cite{IK99} \cite{AR01}, the family
of spacetimes under consideration and the analytic techniques
used have singled out a particular choice of time foliation and
time threading. For the present study of $U(1)$ Symmetric
spacetimes, however, our choice of time is a bit more flexible,
so we may begin to address the issue of the dependence of the
verification of AVTD behavior on the choice of spacetime
observers.

As noted earlier, we use ``harmonic time" in working with the
$U(1)$ Symmetric spacetimes, which means that the time function
$\tau$ satisfies the wave equation $\Box \tau=0$. This condition
does not fix the choice of time foliation; we may freely choose
an inital Cauchy surface, along with an initial choice of lapse
function (This amounts to Cauchy data for the function $\tau$).
The condition $\Box \tau=0$ then determines the rest of the time
foliation.

Fixing the time foliation does not necessarily fix the timelike
observers. They are determined by the the choice of the spacetime
``threading"; ie, the choice of a congruence of timelike paths
(corresponding to the worldlines of the observers). Our analysis
is considerably simplified, however, if we require that the
observer paths be everywhere orthogonal to the leaves of the
spacetime foliation (so the shift vector field is everywhere
zero). With this condition imposed, the choice of spacetime
observers is fixed by the choice of the (harmonic) time function
$\tau$.

While it would be useful to compare the observations relevant to
AVTD behavior that are seen by each set of observers
corresponding to each choice of a harmonic time in a given $U(1)$
Symmetric spacetime, we shall here pursue a more modest goal:
Given a fixed polarized $U(1)$ Symmetric spacetime and a fixed
choice of harmonic time such that the corresponding
surface-orthogonal observers see AVTD behavior, we shall show
that there is a full (two free functions on $T^2$) family of
other harmonic time choices \footnote{We call the parametrization
of this family of time choices by two free functions "full" since
the space of Cauchy data for solutions of $\Box \tau=0$ consists
of two functions.} such that their corresponding
surface-orthogonal observers see AVTD behavior as well. Note that
while we focus on the polarized case here to simplify the
discussion, the same sort of results hold for half-polarized
$U(1)$ Symmetric solutions as well.

We start by identifying the family of alternate harmonic time
choices. To do this, we fix a polarized $U(1)$ Symmetric
spacetime $(T^3\times R, g)$ and a harmonic time choice $\tau$
whose corresponding observers see AVTD behavior. We then write
the harmonic time condition for a new time function $T$; in first
order form, with $\zeta := \partial_{\tau} T$, we have
\begin{eqnarray}
\partial_{\tau}T=\zeta
\label {205}
\end{eqnarray}
\begin{eqnarray}
\partial_{\tau} \zeta
=(e^{-2\tau}e^{\Lambda} e^{ab}T ,_a),_b.
\label {206}
\end{eqnarray}
(Here we are choosing to work with the $2+1$
rather than $3+1$ version of the harmonic time coordinate). Now
noting that
\begin{eqnarray} \hat{T}=a(x)+b(x) \tau
\label {207}
\end{eqnarray}
\begin{eqnarray} \hat {\zeta}=b(x)
\label {208}
\end{eqnarray}
for arbitrary functions $a(x)$ and $b(x)>0$ on
$T^2$ solves the VTD version of equations (\ref{205})-(\ref{206})
(which sets the right hand side of equation (\ref{206}) to zero),
we seek a family of solutions of (\ref{205})-(\ref{206}) of the
form
\begin{eqnarray}
T=a(x)+b(x) \tau +\delta T \label{209}
\end{eqnarray}
\begin{eqnarray}
\zeta=b(x)+e^{-\epsilon \tau} \delta \zeta
\label{210}
\end{eqnarray}
for some $\epsilon>0$. (The condition
$b>0$ is required so that $T$ uniformly approaches infinity as
$\tau$ approaches infinity). Plugging these forms into
(\ref{205})-(\ref{206}), we derive
\begin{eqnarray}
\partial_\tau \delta T = e ^{-\epsilon \tau}
\delta \zeta
\label {211}
\end{eqnarray}
and
\begin{eqnarray}
\partial_\tau \delta \zeta-\epsilon \delta\zeta =
e^{(\epsilon-v_\Lambda+2v_z)\tau}f(\tau,u,v,T,T_{\nabla}),
\label{212}
\end{eqnarray}
where $T_{\nabla}$ denotes a set of new
variables which we introduce so that only first derivatives
appear, and where the function $f$ is analytic in $(u,v,\tau,T,
T_{\nabla})$ and bounded by a polynomial in $\tau$. Thus,
presuming the usual restrictions on $v_\Lambda$ and $v_z$, we
have a Fuchsian system for $\delta T$ and $\delta \zeta$ such
that $T$ and $\zeta$ of the form (\ref{209}) and (\ref{210})
solve (\ref{205})-(\ref{206}). This gives us our family
(parametized by the two functions $a(u,v)$ and $b(u,v)>0)$ of
choices of harmonic time for our fixed spacetime $(T^3\times
R,g)$.

Before addressing the question of observer-dependence of AVTD
behavior for these different foliations, we wish to compare the
paths of the observers orthogonal to the $T=a+b\tau +\delta T$
foliation with those orthogonal to the $\tau$ foliation. One way
to do this is to start with the metric $g(x^i, \tau)$ expressed
in terms of the $\tau$ foliation and the observers $x^i$
orthogonal to the $\tau$ foliation, then re-express $g$ in terms
of the $T$ foliation but with the $\tau$-compatible observers
retained, and finally determine the shift vector $M(x,T)$, which
measures the extent to which the $\tau$-compatible observers fail
to be orthogonal to the $T$ foliation. To carry out this
calculation in practice, what we do is invert the transformation
(\ref{209}), obtaining \footnote{Our assumption that $b>0$ and
our verification that $\delta T$ decays to zero for large $\tau$
guarantees that there exists $c(x)=(\frac{-a(x)}{b(x)}$),
$h(x)=\frac{1}{b(x)}>0$, and $\delta
\tau(x,t)=(\frac{-1}{b(x)} )\delta T$ with $\lim_{T\rightarrow
\infty}\delta \tau=0$ such that (\ref{213}) is indeed the
$x$-parametrized inverse to (\ref{209}).}
\begin{eqnarray}
\tau &=& (-a(x)/b(x))+(1/b(x))T-(1/b(x))\delta
T(x,T) \nonumber \\ &=& c(x)+ h(x)T+ \delta \tau (x,T),
\label{213}
\end{eqnarray}
and then substitute both $\tau(x,T)$
and
\begin{eqnarray}
d\tau=(\partial
_kc+T\partial_kh+\partial_k\delta\tau)dx^k+(h+\frac {\partial
(\delta \tau) }{\partial T})dT
\end{eqnarray}
into the expression (\ref{1}) for $g(x,\tau)$,
which for convenience we rewrite as
\begin{eqnarray}
g(x,\tau)=-n^2(x,\tau)d\tau
^2+\gamma_{ij}dx^idx^j.
\end{eqnarray}
Here the indices $i,j,k$
run from 1 to 3; and $n(x,\tau)$ and $\gamma_{ij}(x,\tau)$ are
certain combinations of $e^\phi$, $e^\Lambda$, etc (see (\ref{1})
along with the discussion just below it). We obtain
\begin{eqnarray}
g(x,T) = -
n^2(h+\partial_T\delta\tau)^2dT^2-2n^2(h+\partial_T\delta\tau)
(\partial_ic+T\partial_ih+\partial_i
\delta \tau)dTdx^i \nonumber \\ +
[\gamma_{ij}-n^2(\partial_ic+T\partial_ih+\partial_i\delta
\tau)(\partial_jc+T\partial_jh+\partial_j\delta\tau)]dx^idx^j\nonumber
\\
\label{214}
\end{eqnarray}
where
$n(x,T)=n(x,c(x)+h(x)T+\delta\tau(x,T))$, etc. If we now compare
expression (\ref{214}) for $g$ with the standard lapse-shift
(``ADM") form of a spacetime metric \begin{eqnarray}
g=-N^2dt^2+h_{ij}(dx^i+M^idt)(dx^j+M^jdt), \end{eqnarray} then we
find that the shift vector $M$ can be expressed as
\begin{eqnarray}
M^i=-\lambda^{ij}[n^2(h+\partial_T\delta\tau)(\partial_jc+
T\partial_jh+\partial_j\delta\tau)]
\label{215}
\end{eqnarray}
where $\lambda^{ij}$ is the inverse of the tensor appearing as
the coefficients of $dx^idx^j$ in (\ref{214}).

To see the behavior of the shift $M^i$ for large values of $T$, we
need to write out n and $\lambda^{ij} $ in terms of the variables
$\Lambda, \phi,x,z$, and then substitute into the asymptotic
expressions for these quantities. In doing so, we again need to
replace $\tau$ by its expression (\ref{213}) in terms of $T$,
noting that such quantities as $\delta\Lambda$ which decay for
large $\tau$ do so as well for large $T$. We find that
\begin{eqnarray}
M^i \approx e^{(2v_z-v_{\Lambda})hT} \times
(\mbox{polynomial in}\, T).
\end{eqnarray}
Thus, recalling our usual restrictions
(\ref{29c})-(\ref{29e}) on $v_\Lambda$ and $v_z$, and also
recalling that $h>0$, we see that the shift decays rapidly to
zero for large $T$. It is especially telling that the shift
decays to zero much more quickly than does the lapse as one
approaches the singularity, since we calculate
\begin{eqnarray}
M^i M_i/N^2 \approx e^{(2v_z-v_{\Lambda)}hT} \times
(\mbox{polynomial in}\, T).
\end{eqnarray}
So we conclude that, for any values of the
harmonic time parameter functions $a(x)$ and $b(x)>0$, the paths
of the surface orthogonal observers corresponding to
$T=a+b\tau+\delta T$ become increasingly parallel to those of the
surface orthogonal observers corresponding to $\tau$.

We now discuss how to verify that the observers which are
orthogonal to the $T=a+b\tau+\delta T$ foliation see AVTD
behavior in the chosen spacetime. The key first step is to find
coordinates $y^k$ which correspond to the $T$ orthogonal
observers. In practice, what we seek is a coordinate
transformation
\begin{eqnarray}
(x,T)\rightarrow(x(y,T),T)
\end{eqnarray}
which
removes the shift term; ie, for which $g$ has no $dydT$ term. A
bit of calculation shows that if we choose the function $x(y,T)$
so that it satisfies
\begin{eqnarray} \partial_T
x^i(y,T)=-M^i(x(y,T),T),
\label{216}
\end{eqnarray}
where $M^i$
is the shift vector field discussed above, then indeed the metric
written in terms of the coordinate $(y,T)$ has vanishing shift,
and so indeed the $y=constant$ observers are $T$ surface
compatible.

We note three features of the coordinate $y$ generated via
(\ref{216}). First, the ODE system (\ref{216}) together with a
set of initial conditions $x^i(y,T_0)=\xi^i(y)$ constitute a
well-posed system, with a unique local solution. Second, although
the system (\ref{216}) is generally nonlinear, it follows from
the boundedness of $M^i$ that (\ref{216}) (together with the
chosen initial conditions) determines a unique solution
$x^i(y,T)$ for all $T$. Third, since $M^i$ decays to zero
exponentially rapidly for large $T$, we may write solutions
$x^i(y,T)$ in the form
\begin{eqnarray} x^i(y,T)=X^i(y)+\delta
x^i(y,T)
\label{217}
\end{eqnarray}
where $\delta x^i(y,T)$
decays to zero for large $T$.

With the transformation between $x$ and $y$ determined, we next
rewrite the metric $g(x,t)$ in terms of the new coordinates. We
are only interested in this for large $T$, so we may use
expression (\ref{217}), >from which we derive \begin{eqnarray}
dx^i \approx X^i,_jdy^j-M^idT. \end{eqnarray} Substituting this
expression together with (\ref{217}) into formula (\ref{214}),
and also replacing $n$ and $\gamma_{ij}$ by the appropriate
expressions in terms of ${\cal A}_{pol}=\{\phi_o, \Lambda_o, x_o,
z_o, v_\phi, v_\Lambda, v_x, v_z \}$ and $T$, we obtain a
coordinate representation of the metric of the form
\begin{eqnarray} g(y,T)=g_{TT}(y,T)dT^2+g_{ij}(y,T)dy^idy^j,
\label{218} \end{eqnarray} with $g_{TT}$ and $g_{ij}$ as fairly
complicated functions involving ${\cal A}_{pol}$ along with the
functions $c$, $h$, and $X^i(y)$.

To show that the $T$ surface orthogonal observers see AVTD
behavior, it is sufficient to verify the following:
\begin{proposition}
For every choice of the asymptotic data
${\cal A}_{pol}$ and for every choice of the harmonic time
transformation functions $\{a,b > 0\}$, there exist functions
$\{\tilde \phi_(y), \tilde\Lambda_o(y), \tilde x_o(y), \tilde
z_o(y), \tilde v_\phi(y), \tilde v_\Lambda(y), \tilde v_x(y),
\tilde v_z(y)\}$, and functions $\{\delta \tilde \phi(y,T),
\delta \tilde \Lambda(y,T),\delta \tilde x(y,T), \delta \tilde
z(y,T),\delta \tilde p_\phi (y,T), \delta \tilde p_\Lambda (y,T),
\delta \tilde p_x (y,T), \delta \tilde p_z(y,T)\}$ decaying to
zero for large $T$, such that the metric coefficients $g_{TT}$
and $g_{ij}$ in (\ref{218}) can be written in the polarized
$U(1)$ Symmetric AVTD form described in sections 2 and 3; in
particular, one has the asymptotic behaviors
\begin {eqnarray}
g_{TT} \approx -h^2 e^{2(\Lambda-2 \tau)}
e^{-2\phi}
\label {219}
\end {eqnarray}
\begin {eqnarray}
g_{ij} \approx \gamma_{mn} \frac{\partial
X^m}{\partial y^i} \frac{\partial X^n}{\partial y^j}
\label{219a}
\end{eqnarray}
\end {proposition}

The proof of this proposition is a straightforward (somewhat
tedious) consequence of carrying through the details of
calculation (\ref{218}). We omit it here, together with the
explicit (not especially useful) expressions one obtains for
$\{\tilde \phi_(y), \tilde\Lambda_o(y), \tilde x_o(y), \tilde
z_o(y), \tilde v_\phi(y), \tilde v_\Lambda(y), \tilde v_x(y),
\tilde v_z(y)\}$ in terms of ${\cal A}_{pol}$ and $\{a,b>0\}$.

Do we see AVTD behavior in these spacetimes using other
foliations and sets of observers? This is not yet known. The
foliations and observers we have discussed here exhibit two
important features asymptotically: 1) All of the sets of surface
compatible observers become parallel as one approaches the
singularity. 2) In all of them, one sees AVTD behavior. It is not
clear whether these features are related or not. It may be, for
example, that in a spacetime which shows AVTD behavior with
respect to one foliation and set of observers, AVTD behavior is
seen by the surface compatible observers of any other foliation
if and only if those observers become parallel asymptotically to
the original ones. Or, it might be the case that these two
features which coincide for our harmonic foliations and observers
in these $U(1)$ Symmetric solutions are not closely related more
generally. In either case, it may be that the surface compatible
observers corresponding to most if not all spacelike foliations
in a spacetime with AVTD behavior are asymptotically parallel, as
one approaches the singularity. We hope to explore this issue in
future work.

\section{Generating Further U(1) Symmetric Spacetimes with AVTD
Behavior}

Numerical work \cite{BM98b} suggests that general $U(1)$
Symmetric vacuum solutions on $T^3 \times R$ show Mixmaster
behavior near the singularity. We may then ask if the $U(1)$
Symmetric solutions with AVTD behavior extends beyond the classes
we have discussed here so far. In this section, we use Geroch's
$SL(2,R)$ method of generating new $U(1)$ Symmetric solutions
from old ones to show that this is the case.

We recall \cite{Ge71} \cite{GM94} how the Geroch transformation
works: If $\{
\phi,
\Lambda, w, x, z;\- p_\phi, p_\Lambda, r, p_x. p_z\}$ is a
solution of the $U(1)$ Symmetric vacuum Einstein equations on
$(\Sigma^3,R)$, and if $\left( \begin{array}{clcr} a& b\\c& d
\end{array} \right)$ is a (constant) matrix contained in
$SL(2,R)$ ( so det$\left( \begin{array}{clcr} a& b\\c& d
\end{array} \right)=1$), then $\{\hat \phi, \Lambda, \hat w, x,
z; \hat p_\phi, p_\Lambda, \hat r, p_x. p_z\}$ with \begin
{eqnarray} e^{2\hat \phi}= {e^{2\phi}
\over{c^2(w^2+e^{4\phi})+2cdw+d^2}} \end {eqnarray} \begin
{eqnarray} \hat w={{ac(w^2+e^{4\phi})+(ad+bc)w+bd}\over
{c^2(w^2+e^{4\phi}) +2cdw+d^2}} \end{eqnarray} \begin{eqnarray}
\hat
p_\phi={{p(c^2(w^2-e^{4\phi})+2cdw+d^2)-r(4e^{4\phi}(cd+wc^2))}
\over {c^2(w^2+e^{4\phi})+2cdw +d^2}}
\end{eqnarray}
\begin{eqnarray} \hat r=p (c^2w+cd) +r[d^2 + c^2(w^2-e^{4\phi})
+2cdw]
\end{eqnarray} is also a solution of the $U(1)$ Symmetric vacuum
Einstein equations on $\hat \Sigma^3 \times R$, where $\hat
\Sigma^3$ is diffeomorphic to
$\Sigma^3$ if and only if $\int_{\hat \Sigma^3} \hat r =
\int_{\Sigma^3}r$.

We wish to consider Geroch transformations which map solutions on
$\Sigma^3=T^3$ to others also on $T^3$. Since
$\int_{\Sigma^3}r=0$ if and only if $\Sigma^3=T^3$, we seek
solutions and transformations such that $\int_{\Sigma^3}r=0$ both
before and after the the transformation. Using the three
constants of the motion
\begin{eqnarray} A=\int_{\Sigma^3}(2wr+p_\phi), \end{eqnarray}
\begin{eqnarray} B=\int_{\Sigma^3}r,
\end{eqnarray} and
\begin{eqnarray} C=\int_{\Sigma^3}[r(e^{4\phi}-w^2)-p_\phi w],
\end{eqnarray} one finds that if \{A,B,C\} characterize the
original solution, then \begin{eqnarray} \hat
A=(ad+bc)A+2bdB-2acC,
\end{eqnarray}
\begin{eqnarray}
\hat B=d^2B-c^2C+cdA,
\end{eqnarray} and
\begin{eqnarray}
\hat C=a^2C-b^2B-abA
\end{eqnarray} characterize the transformed solution. So we seek
solutions and $SL(2,R)$ matrices $\left( \begin{array}{clcr} a&
b\\c& d \end{array} \right)$ such that $B=0$ and
$d^2B-c^2C+cdA=0.$

For polarized solutions, $r=0=w$ and $p_\phi\neq 0$, so $B=0$,
$C=0$, and $A \neq 0$. Thus, to obtain $\hat B=0$, we need to
choose $\left( \begin{array}{clcr} a& b\\c& d \end{array}
\right)$ with either $c=0$ or $d=0$ (we can not have both zero,
since $\left( \begin{array}{clcr} a& b\\c& d \end{array} \right)
\in SL(2,R).$ ). For half-polarized solutions as well, the
asymptotic decay of $w$ and $e^{4\phi}$ require that $C=0$, so
again we obtain $\hat B=0$ if and only if either $c=0$ or $d=0$.

Applying an $SL(2,R)$ Geroch transformation with $c=0$ to either
a polarized or half-polarized solution produces nothing really
new. \footnote {Such a transformation results in a (constant)
rescaling of some of the quantities, and the addition of a
constant to $w$. We recall that $w$ does not appear in the
metric, and it appears in the Hamiltonian and in the field
equations only as a gradient term $\nabla w$.} The $d=0$
transformation, however, produces geometrically new AVTD
solutions. Applied to a polarized solution, such a transformation
takes the form
\begin{eqnarray}
e^{2\hat \phi}=(1/c^2)e^{-2
\phi}
\end{eqnarray}
\begin {eqnarray}
\hat w=a/c,
\end{eqnarray}
\begin{eqnarray}
\hat p_\phi=-p_ \phi
\end{eqnarray}
and
\begin {eqnarray}
\hat r=0.
\end{eqnarray}
The new solutions are geometrically distinct from
the original ones, since $\int e^{2 \phi}dx^3$ is the diameter of
the (evolving) three-geometry along the $U(1)$ symmetry
direction; and while this diameter decays to zero in the original
solution, it blows up in the transformed solution. Note that one
readily verifies that so long as the original solutions are AVTD,
the transformed one are as well.

For the half-polarized solutions, the transformation
corresponding to $\left( \begin{array}{clcr} a& b\\c& d
\end{array} \right)$ takes the form (with ''$\rightarrow"$
indicating asymptotic values)
\begin {eqnarray} e^{2 \hat \phi} =
{e^{2 \phi}\over{c^2(w^2+e^{4 \phi})}}\nonumber \\
\rightarrow
e^{-2 \phi}/c^2
\end{eqnarray}
\begin {eqnarray}
\hat w= {{ac(w^2+e^{4 \phi}) +bcw} \over
{c^2(w^2+e^{4 \phi})}}
\nonumber \\
 \rightarrow {\left(a \over c\right)}+ {\left( b \over c
\right)}{\left(-\stackrel{\circ} r \over 4v_\phi
\right)}
\end{eqnarray}
\begin {eqnarray}
\hat p_\phi
={{p_\phi(c^2(w^2-e^{4\phi}))-r(4e^{4\phi}c^2w)} \over
{c^2(w^2+e^{4\phi})}} \nonumber \\
 \rightarrow -p_{\phi}
\end{eqnarray}
\begin{eqnarray}
\hat r=p(c^2w)+rc^2(w^2-e^{4\phi}) \\ \nonumber \rightarrow 0.
\end{eqnarray} Here again, the inversion of $e^{2\phi}$ indicates
geometrically new AVTD solutions. As well, while the
half-polarized solutions discussed above have $w$ asymptotically
vanishing and $r$ approaching a general function on $T^2$ (see
equations (\ref{30a})-(\ref{30b})), these new solutions have $r$
asymptotically vanishing and $w$ approaching a general function
on $T^2$. Examining the Hamiltonian (\ref{6a})-(\ref{6b}), we see
that this swapping of the asymptotic behavior of $r$ and $w$ is
needed to avoid having $H$ blow up as a consequence of the
inversion of $e^{2 \phi}$.

\section {Conclusion}

How prevalent is AVTD behavior in solutions of Einstein's
equations? Since this behavior is so specialized, it is perhaps
rather surprising that, as we show here, it is found in
substantial classes of solutions with only one Killing field. As
noted above, numerical work \cite{BM98b} indicates that in
generic cosmological spacetimes with one Killing field, one finds
Mixmaster rather than AVTD behavior near the singularity.
However, it is likely that we can extend the classes of solutions
known to have AVTD behavior in at least two ways.

First, we should be able to remove the analytic condition which
our theorems here require. Rendall has shown how to do this for
the class of Gowdy spacetime: In his work with Kichenassamy
\cite{KR98}, it is shown that certain classes of analytic Gowdy
spacetimes have AVTD behavior; then in his later work \cite{Re00}
, he shows that the same holds for $C^{\infty}$ solutions. Work
hs begun which applies techniques similar to those used in
\cite{Re00} to show that AVTD behavior is found in classes of
$C^{\infty}$ polarized
$U(1)$ Symmetric spacetimes.

Second, we hope to be able to show that all of the polarized
$U(1)$ Symmetric vacuum solutions have AVTD behavior, and not
just a subset of these spacetimes. Numerical studies suggest that
this is true; we are looking for ways to prove this contention.

It is not clear whether there are vacuum solutions with
$\it{no}$ Killing fields which exhibit AVTD behavior. Andersson
and Rendall \cite{AR01} have shown that there are classes of
solutions of the Einstein equations coupled to a scalar field
with AVTD behavior; however, while heuristic analyses have
predicted this result, the same sorts of studies suggest that
generic vacuum solutions with no Killing fields should show
Mixmaster rather than AVTD behavior. We note that, to date, apart
from the work of Berger and Moncrief \cite{BM00} which uses
Geroch transformations to generate a small class of $U(1)$
Symmetric spacetimes with Mixmaster behavior, there are very few
rigorous results concerning the existence of such behavior in
spatially inhomogeneous cosmological spacetimes.

\section {Acknowledgments}

We would like to thank the Institute for Theoretical Physics in
Santa Barbara, the Albert Einstein Institute in Golm, the Erwin
Schroedinger Institute in Vienna, and L'Institute pour les Hautes
Etudes Scientifiques at Bures-sur-Yvette for providing very
pleasant, supportive, and stimulating environments for our
collaboration on this work. This work was partially supported by
the NSF, under grants PHY-9800732 and PHY-0099373 at Oregon and
grants PHY-9732629 and PHY-0098084 at Yale.

\end{document}